\documentclass[aps,prl,preprint,11pt,superscriptaddress,bibnotes,amsmath,amssymb,amsfonts,showkeys]{revtex4-2}

\usepackage{graphicx}
\usepackage{dcolumn}
\usepackage{bm}
\usepackage{multirow}
\usepackage{float}
\newcolumntype{C}[1]{>{\centering\arraybackslash}p{#1}}

\begin{document}

\title{Strong Edge Stress in Molecularly Thin Organic--Inorganic Hybrid Ruddlesden--Popper Perovskites and Modulations of Their Edge Electronic Properties}

\author{Devesh R. Kripalani}
\affiliation{School of Mechanical and Aerospace Engineering, Nanyang Technological University, 50 Nanyang Avenue, Singapore 639798, Singapore}

\author{Yongqing Cai}
\email[\textbf{Corresponding Authors:}\\E-mail: ]{yongqingcai@um.edu.mo (Yongqing Cai); kzhou@ntu.edu.sg (Kun Zhou)\\}
\affiliation{Joint Key Laboratory of the Ministry of Education, Institute of Applied Physics and Materials Engineering, University of Macau, Avenida da Universidade, Taipa, Macau 999078, P. R. China}

\author{Jun Lou}
\affiliation{Department of Materials Science and NanoEngineering, Rice University, Houston, Texas 77005, USA}

\author{Kun Zhou}
\email[\textbf{Corresponding Authors:}\\E-mail: ]{yongqingcai@um.edu.mo (Yongqing Cai); kzhou@ntu.edu.sg (Kun Zhou)\\}
\affiliation{School of Mechanical and Aerospace Engineering, Nanyang Technological University, 50 Nanyang Avenue, Singapore 639798, Singapore}
\affiliation{Environmental Process Modelling Centre, Nanyang Environment and Water Research Institute, Nanyang Technological University, 1 Cleantech Loop, Singapore 637141, Singapore}


\begin{abstract}
Organic--inorganic hybrid Ruddlesden--Popper perovskites (HRPPs) have gained much attention for optoelectronic applications due to their high moisture resistance, good processibility under ambient conditions, and long functional lifetimes. Recent success in isolating molecularly thin hybrid perovskite nanosheets and their intriguing edge phenomena have raised the need for understanding the role of edges and the properties that dictate their fundamental behaviours. In this work, we perform a prototypical study on the edge effects in ultrathin hybrid perovskites by considering monolayer (BA)$_2$PbI$_4$ as a representative system. Based on first-principles simulations of nanoribbon models, we show that in addition to significant distortions of the octahedra network at the edges, strong edge stresses are also present in the material. Structural instabilities that arise from the edge stress could drive the relaxation process and dominate the morphological response of edges in practice. A clear downward shift of the bands at the narrower ribbons, as indicative of the edge effect, facilitates the separation of photo-excited carriers (electrons move towards the edge and holes move towards the interior part of the nanosheet). Moreover, the desorption energy of the organic molecule can also be much lower at the free edges, making it easier for functionalization and/or substitution events to take place. The findings reported in this work elucidate the underlying mechanisms responsible for edge states in HRPPs and will be important in guiding the rational design and development of high-performance layer--edge devices.
\end{abstract}


\maketitle

\setcitestyle{super}

\section{Introduction}
\begin{figure*}
\includegraphics[width=1\textwidth]{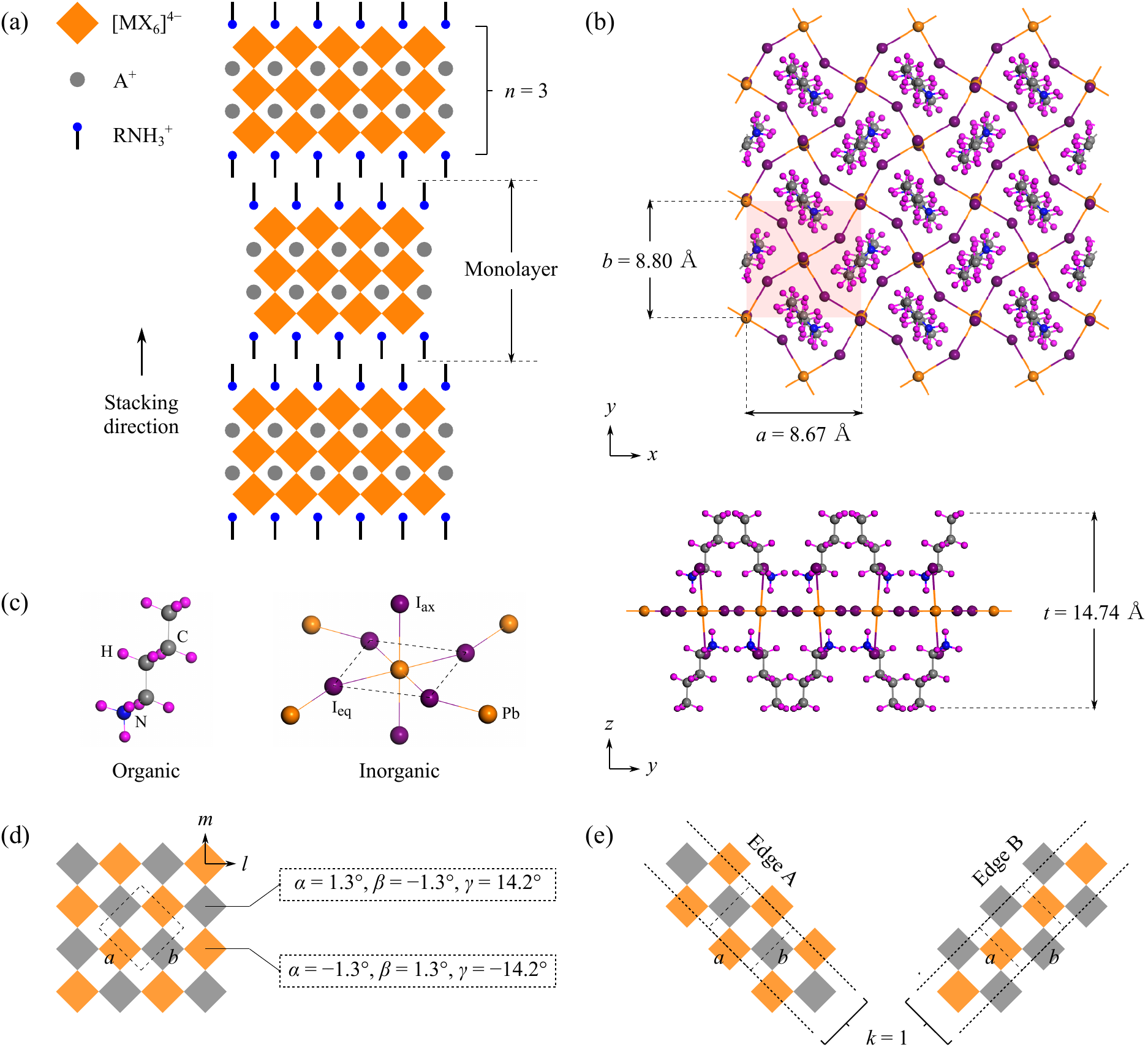}
\caption{(a) Schematic of the HRPP crystal structure along the stacking direction. In this example, the inorganic framework consists of three layers of [MX$_6$]$^{4-}$ octahedra (\textit{i.e.,} $n = 3$). (b) Crystal structure of monolayer (BA)$_2$PbI$_4$ (C, grey; H, pink; N, blue; Pb, orange; I, purple), with the orthorhombic unit cell (shaded pink) indicated in the top ($xy$) view. (c) Atomic configurations of the organic BA molecule and inorganic [PbI$_6$]$^{4-}$ octahedron that constitute the (BA)$_2$PbI$_4$ perovskite. (d) Schematic of the octahedra tilt system (top view) in monolayer (BA)$_2$PbI$_4$, with the orthorhombic unit cell indicated by dashed lines. (e) Construction of (BA)$_2$PbI$_4$ nanoribbons with identical edges along the two fundamental lattice directions. \label{HRPP_config}}
\end{figure*}

Organic--inorganic hybrid Ruddlesden--Popper perovskites (HRPPs) have recently emerged to great scientific interest as exciting two-dimensional (2D) layered materials for next-generation solid-state optoelectronics.\cite{MHhrpp16,TNhrpp16,CShrpp15,SHhrpp14,KMhrpp99} This class of compounds have a general formula of (RNH$_3)_2$A$_{n-1}$M$_n$ X$_{3n+1}$, where RNH$_3$ is an aliphatic or aromatic alkylammonium organic cation spacer (\textit{e.g.,} butylammonium (BA), phenylethylammonium (PEA)), A is a monovalent cation, in most cases methylammonium (MA), M is a divalent metal (\textit{e.g.,} Ge$^{2+}$, Pb$^{2+}$), and X is a halide (\textit{e.g.,} Cl$^-$, I$^-$). HRPPs typically consist of an inorganic $n$-layer network of vertex-linked [MX$_6$]$^{4-}$ octahedra stabilized by A$^+$ groups in its cavities and sandwiched between two organic buffer layers containing RNH$_3$\textsuperscript{+} spacers, as shown in Fig. \ref{HRPP_config}(a). As a result, this gives rise to a periodic quantum well-like heterostructure in which the semiconducting inorganic framework serves as the potential well and the insulating organic layers as potential barriers.\cite{MTmqw95,HImqw92} The index $n$ is commonly referred to as the HRPP's dimensionality, and it presents a particularly useful route for tuning the material's electronic and optical characteristics through quasi-2D $(1 < n < \infty)$ or 2D $(n = 1)$ well confinement. For instance, thicker inorganic slabs (higher $n$) have been shown to reduce the band gap and exciton binding energy, and can lead to solar cell devices with improved power conversion efficiencies (PCEs).\cite{BAqwell18,GBqwell18,XMqwell17,QYqwell16,SCqwell16}

HRPPs have risen as promising alternatives over traditional three-dimensional (3D) hybrid perovskites owing to their superior moisture resistance, better processibility under ambient conditions, and longer functional lifetimes.\cite{TNhrpp16,CShrpp15,SHhrpp14,QYqwell16,SHstab19} The organic spacer molecules, which are hydrophobic, play a central role in maintaining the air, heat and light stability of the perovskite.\cite{SHstab19} These molecules enforce the 2D-layered nature of HRPPs, distinguishing them fundamentally from their 3D-bonded predecessors. It is widely known that the structures of 3D hybrid perovskites, most prominently MAPbX$_3$ (where X = Cl, Br, I), can be unstable to temperature, and transitions can occur across cubic, tetragonal and orthorhombic phases.\cite{OM3dphase90,PW3dphase87} Moreover, the planar surfaces of HRPPs enable effective encapsulation and the formation of van der Waals heterojunctions with other 2D materials (graphene, h-BN, MoS$_2$, \textit{etc.}).\cite{FWplanar19,LAplanar18,SGplanar18} Although lead halide perovskites remain at the forefront of most studies, the successful synthesis of lead-free HRPPs based on tin,\cite{FWplanar19,CSpbfree17} germanium \cite{CWpbfree17} and even binary metal combinations \cite{MMpbfree20} like Ag--Bi and Cu--In has motivated a push for more eco-friendly candidates.

The 2D-layered structure of the Ruddlesden--Popper phase allows for the facile isolation of molecularly thin layers ($<$10 nm) \textit{via} mechanical exfoliation.\cite{GWexf16,YCexf15,NEexf14} Direct bottom-up growth of single- and few-layer HRPP nanosheets has also been reported from solution-phase deposition on Si/SiO$_2$ substrates.\cite{GSsolv19,DWsolv15} Such ultrathin hybrid perovskites are somewhat intriguing as their mechanical, electronic and optical properties can be vastly different from those in bulk.\cite{LAplanar18,DWsolv15,TSultra2d18,QZultra2d18} Previous works on low-dimensional perovskites have suggested the importance of surface doping \cite{OZlowdp18} and ligand engineering \cite{CZlowdp19} as ways to improve their performance and stability. For example, Chen \textit{et al.} have shown that by optimizing the composition of surface ligands on $\alpha$-CsPbI$_3$ quantum dots, the PCE of photovoltaic devices can be enhanced from 7.8\% to 11.9\%.\cite{CZlowdp19} Nevertheless, research involving molecularly thin HRPPs is still in its infancy, and further exploration of the single quantum well configuration is necessary to fully harness the physics associated with extreme thickness.

Meanwhile, the edges of HRPPs have come under intense scrutiny in recent years. In 2017, Blancon \textit{et al.} discovered that the edges of (BA)$_2$(MA)$_{n-1}$Pb$_n$I$_{3n+1}$ can play host to lower-energy states---termed as layer--edge states (LESs) in their work---that are capable of dissociating excitons into longer-lived free carriers, leading to significantly higher photovoltaic cell efficiences.\cite{BTles17} However, the identity of these states was not clear. Subsequent experiments went on to suggest that such LESs were in fact composed of a self-formed 2D/3D perovskite heterostructure following the loss of BA ligands at the free edges,\cite{QDhetedge20} and reversible control over its formation could be achieved through BAI and MAI treatment.\cite{ZTrevcon19} It was also shown that LES formation is not entirely an intrinsic process but rather, one that could be facilitated under humid conditions.\cite{SDhumid19} Despite these efforts, a fundamental theoretical basis for predicting and controlling the behaviour of edges is lacking. Besides, localized edge effects can be especially important in low-dimensional systems; edges have been found to strongly influence the structure, electronic properties, magnetism and catalytic activity of many classical 2D materials and their heterostructures, including those of graphene \cite{GYedgegrtmd17,HLedgegr09,SRedgegr08} and MoS$_2$.\cite{GYedgegrtmd17,SPedgetmd20,CSedgetmd15} Therefore, an atomistic understanding of layer--edge interactions in molecularly thin HRPPs is critical for the development of high-performance perovskites required for light detection \cite{FWplanar19,LAplanar18} and molecular sensing \cite{CLgasdet18} applications.

In this article, our focus lies on the widely familiar (BA)$_2$(MA)$_{n-1}$Pb$_n$I$_{3n+1}$ hybrid perovskite system. More specifically, by considering the $n = 1$ monolayer, we investigate the mechanics and electronic properties of edges from first-principles calculations by exploring a series of well-calibrated nanoribbon models. The nanoribbons are designed to resemble the natural termination of crystal periodicity at the perovskite edges, and as virtual 1D systems, they provide a computationally robust and tractable way to access the edge structure.\cite{ZFedgesim19} Our simulations reveal the existence of significant edge stresses in (BA)$_2$PbI$_4$ that could drive the relaxation process and dominate the morphological response of edges in practice. Furthermore, we find that there is a greater tendency for molecular desorption to occur at the edges as compared to the internal region of the monolayer, thus enhancing the chemical activity of these peripheral areas. The study presented herein puts forth the key metrics necessary for understanding the physics that underpin the behaviour of edges in HRPPs, which, if controlled, can be used to improve the optoelectronic and sensing performance of this emerging class of 2D materials.

\section{Results and discussion}
In this work, monolayer (BA)$_2$PbI$_4$ is taken as a prototypical system for investigating edge effects in molecularly thin HRPPs. Density functional theory (DFT) simulations are performed to model this 2D material at the atomic level and to characterize its mechanical and electronic properties due to the formation of edges (details of our computational procedure are provided in the Methods section). It has been experimentally confirmed that (BA)$_2$PbI$_4$ crystallizes in the primitive centrosymmetric orthorhombic space group $Pbca$ with bulk lattice constants of $a$ = 8.68--8.69 \AA, $b$ = 8.86--8.88 \AA\ and $c$ = 27.57--27.63 \AA.\cite{LAplanar18,BLcryst07,Mcryst96} Accordingly, the crystal structure of monolayer (BA)$_2$PbI$_4$ is modelled as shown in Fig. \ref{HRPP_config}(b). Its lattice constants are calculated to be $a$ = 8.67 \AA\ and $b$ = 8.80 \AA, which are comparable to those of the bulk structure. This finding is indicative of the weak interlayer coupling within the material, and is corroborated by Leng \textit{et al.} who have also reported a negligible effect of the bulk-to-monolayer transition on the lattice parameters.\cite{LAplanar18}

Figure \ref{HRPP_config}(c) illustrates the atomic structures of the organic and inorganic components that make up the hybrid perovskite. The organic BA molecule (chemical formula: CH$_3$(CH$_2$)$_3$NH$_3$\textsuperscript{+}) has an alkyl chain length of four carbon atoms and is oriented along the normal ($z$) direction with its ammonium head pointing towards the anionic inorganic framework. The inorganic framework in turn consists of an interconnected single-layer network of [PbI$_6$]$^{4-}$ octahedra and propagates across the 2D ($xy$) plane. Each octahedron is defined by a central Pb atom surrounded by two axial (I\textsubscript{ax}) and four equatorial (I\textsubscript{eq}) iodine atoms. Interactions between the organic and inorganic components, primarily through electrostatic attraction and hydrogen bonding mechanisms, cause the octahedra to adopt tilted configurations. The equilibrium bond lengths Pb--I\textsubscript{ax} and Pb--I\textsubscript{eq} are found to be 3.25 \AA\ and 3.19 \AA, respectively, while the angle of connectivity between adjacent octahedra $\angle$Pb--I\textsubscript{eq}--Pb is 150.7$^\circ$. In general, factors such as the dimensionality ($n$) of the hybrid perovskite and the chemical composition of its organic and inorganic parts can have an influence on the extent of octahedral tilting.\cite{SCqwell16,SHstab19,Mcryst96,SNtilt19} To formally classify the tilt system in the monolayer, an analysis of the tilt pattern is conducted in accordance with the Glazer framework.\cite{Gtilt72} Here, the orientation of each octahedron in the unit cell is evaluated with respect to the pseudocubic axes ($lmn$) before tilting based on the rotation matrix $\textbf{R} = R_{l}(\alpha) R_{m}(\beta) R_{n}(\gamma)$, where $\alpha$, $\beta$ and $\gamma$ denote the tilt angles about each axis.

A necessary criterion for preserving the structural integrity of (BA)$_2$PbI$_4$ is that a tilt of a particular octahedron about a certain axis must be counteracted by equal and opposite tilts of its nearest-neighbor octahedra about that same axis. This tilt system, which we represent schematically in Fig. \ref{HRPP_config}(d), corresponds to an octahedra arrangement that is purely out-of-phase (\textit{i.e.,} $l^{^{-}} m^{^{-}}$ in Glazer notation). Evidently, octahedral tilting is dominant ($\sim$14$^\circ$) about the out-of-plane axis, whereas almost no tilting occurs about the in-plane axes (tilt angles are near 0$^\circ$). At edges, the break in crystal periodicity and abrupt disruption of the tilt system can lead to significant structural distortion and alter the inherent properties of the monolayer. Edge formation along the two fundamental lattice directions is preliminarily considered. As shown in Fig. \ref{HRPP_config}(e), nanoribbon models containing identical edges are constructed, where edge A (B) is defined as that aligned along the $x$ ($y$) direction (see Fig. \ref{HRPP_config}(b)), and the width index $k$ refers to the edge-to-edge span of the nanoribbon normalized with respect to the lattice constant $b$ ($a$). It is noted that nanoribbon simulations of $k > 3$ can however become computationally prohibitive. Fortunately, as we will show later, such large models are not required to capture the edge behaviors of monolayer (BA)$_2$PbI$_4$. In particular, the I-terminated edge structure has been chosen for this investigation as unsaturated iodine bonds have been found to bear strong charge localization effects, which can be important for driving charge separation at the edges.\cite{ZFedgesim19}

The edge energy is calculated as $U_{\text{edge}} = (U_{\text{ribbon}} - pU_{\text{f.u.}})/(2L)$, where $U_{\text{ribbon}}$ is the total energy of the nanoribbon, $U_{\text{f.u.}}$ is the energy per formula unit in a perfect 2D sheet, $p$ is the number of formula units used to model the nanoribbon, and $L$ is the edge length. For each unit cell of the nanoribbon, $p$ is related to the width index $k$ by $p = 2k + 1$. From our analysis of $k = 1, 2, 3$ nanoribbons, we find that $U_{\text{edge}}$ is $\sim$0.1 eV/\AA\ for both edges A and B. The comparable energy costs for forming edges A and B are in agreement with the fact that the bonding configurations are indeed similar along each of the fundamental lattice directions of the (BA)$_2$PbI$_4$ monolayer. To clarify this aspect from a mechanical standpoint, the 2D elastic constants of the monolayer are derived following the formulation described in the Supporting Information (SI). We determine the Young's modulus $E$ to be almost identical along either lattice direction ($E_x$ = 11.9 GPa, $E_y$ = 11.5 GPa), thereby affirming the near-isotropic nature of the (BA)$_2$PbI$_4$ structure. As such, for the remainder of this study, we will turn our focus simply to the case of edge A structures with the knowledge that these results are representative of the response of edge B as well.

\begin{figure*}
\includegraphics[width=1\textwidth]{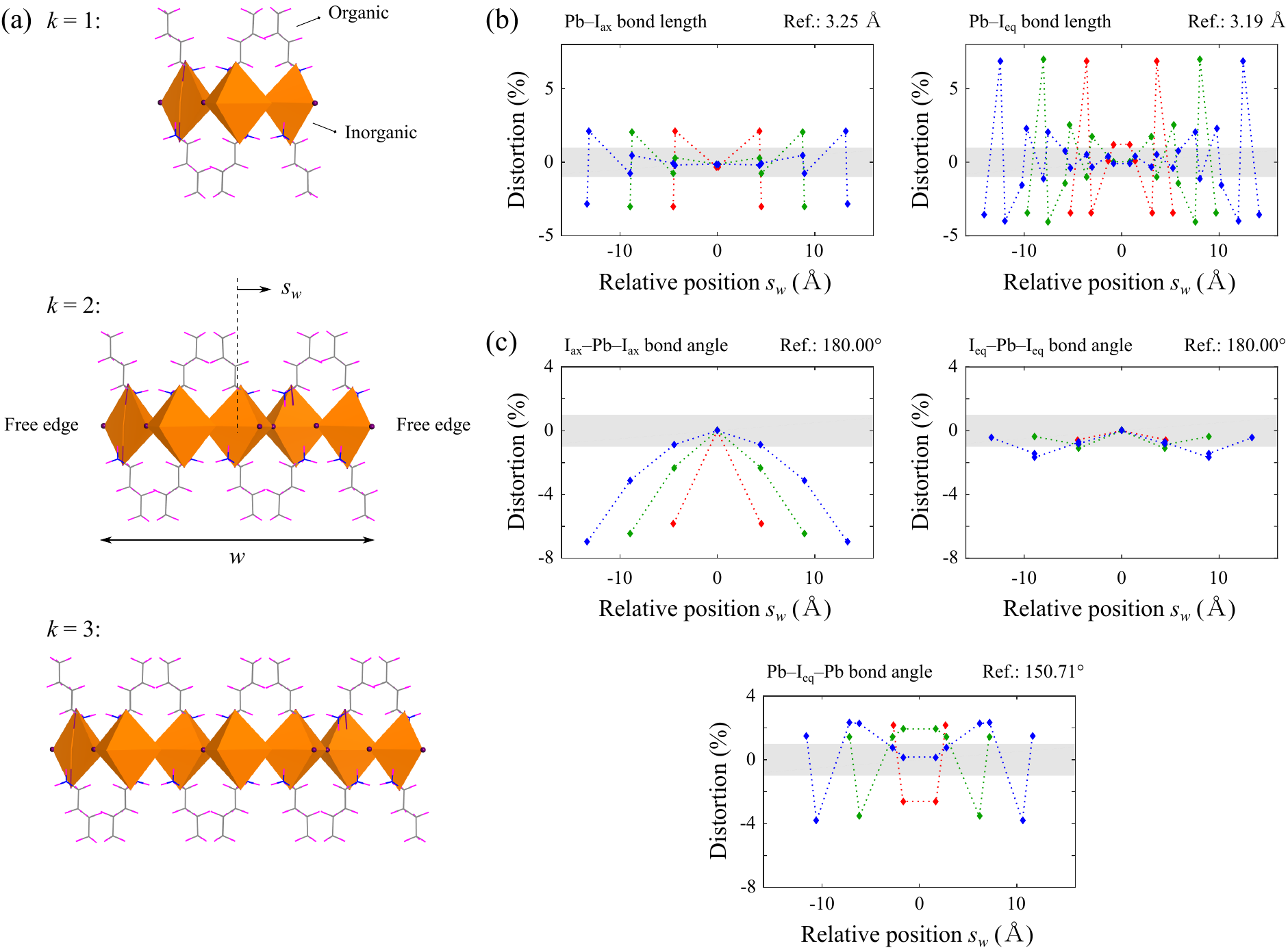}
\caption{(a) Relaxed structures of monolayer (BA)$_2$PbI$_4$ nanoribbons (side view) of width index $k$. Edge-induced distortions in the (b) bond lengths and (c) bond angles across $k$ = 1 (red), $k$ = 2 (green), and $k$ = 3 (blue) nanoribbons. The geometry of the 2D sheet is taken as reference and the grey-shaded regions denote the $\pm$1\% window of fluctuation. \label{1DHRPP_structure}}
\end{figure*}

The relaxed structures of the edge A nanoribbons ($k = 1, 2, 3$) are shown in Fig. \ref{1DHRPP_structure}(a). Figures \ref{1DHRPP_structure}(b) and \ref{1DHRPP_structure}(c) chart the variations in their bond lengths and angles, highlighting the structural distortions brought about in the vicinity of edges. Note that the geometry of the 2D sheet is taken as reference for this analysis. All nanoribbon structures exhibit similar deformation features. At the free edge, the Pb--I\textsubscript{eq} bond length can increase beyond 5\%, while the I\textsubscript{ax}--Pb--I\textsubscript{ax} bond angle can decrease by more than 6\%. Along with the fluctuations observed in the Pb--I\textsubscript{eq}--Pb bond angle, these geometrical changes are reflective of the large shape distortions of the [PbI$_6$]$^{4-}$ octahedra located near the edges. Interactions between the two edges are highly significant for $k = 1$, but as the width of the nanoribbon is increased to $k = 3$, edge--edge interactions gradually dissipate and the central region of the structure begins to retain its 2D sheet-like geometry. The edge-induced structural changes combined with edge--edge interactions can be responsible for modulating the electronic properties of the perovskite. This issue is addressed in greater detail in the later part of this section. First, the mechanical properties of the edge structure, including the edge elastic modulus and edge stress, is characterized.

\begin{figure*}
\includegraphics[width=1\textwidth]{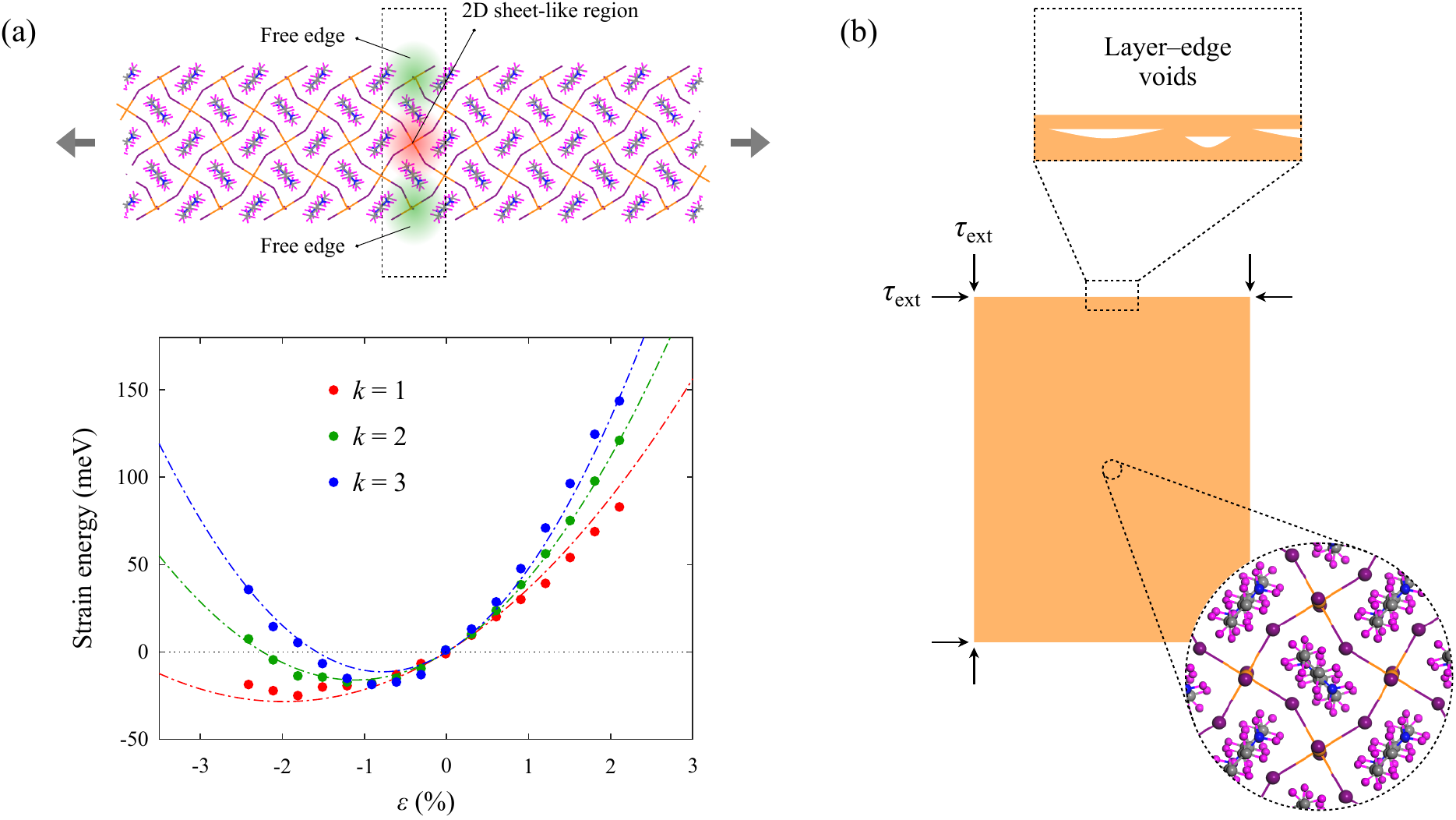}
\caption{(a) Upper panel: Schematic of a (BA)$_2$PbI$_4$ nanoribbon (top view) under axial strain. The rectangle marks one unit cell of the structure which contains the free edges (green) and a central 2D sheet-like region (red). Lower panel: Evolution of the strain energy with strain. The dash-dotted lines denote the respective fits of the data to the analytical relation defined by Eq. (\ref{Us}). (b) Compressive forces $\tau_{\text{ext}}$ exerted on a planar sheet by edges with tensile edge stresses. The insets illustrate the 2D structure of (BA)$_2$PbI$_4$, and voids at the layer--edge region. \label{1DHRPP_energetics}}
\end{figure*}

Let us consider a nanoribbon subject to a state of uniaxial strain $\epsilon$ along the edge direction, as shown in Fig. \ref{1DHRPP_energetics}(a) (upper panel). The strain energy $U$\textsubscript{s} stored in the nanoribbon is given by

{\begin{align}\label{Us}
U\textsubscript{s} (\epsilon, k) = \frac{1}{2}\phi_{k}E\textsubscript{sheet}t_{k}w_{k}L_{0}(1+\epsilon)\epsilon^{2} + E_{\text{edge}}t_{k}L_{0}(1+\epsilon)\epsilon^{2} + 2t_{k}L_{0}f_{\text{edge}}\epsilon
\end{align}}

where $L_{0}$ is the equilibrium lattice constant of the monolayer, $t_{k}$ and $w_{k}$ are the thickness and width of the nanoribbon of index $k$, $f_{\text{edge}}$ refers to the edge stress, $E\textsubscript{sheet}$ and $E_{\text{edge}}$ are the elastic moduli of the 2D sheet and edge, respectively, and $\phi$ is a factor of mixity between 0 and 1 which represents the degree of development of the central 2D sheet-like region within the nanoribbon. A mixity factor of $\phi$ = 1 denotes that a 2D sheet-like region has fully developed and there are no interactions between the edges. Conversely, intermediate values of $\phi$ in ultranarrow nanoribbons indicate that edge--edge interactions are significant in these structures. It is to be noted that the second and third terms in Eq. (\ref{Us}) contain factors of 2 in order to account for the contributions from the two edges of the nanoribbon.

The edge properties, $E_{\text{edge}}$ and $f_{\text{edge}}$, are determined by computing the energies of strained nanoribbon configurations and fitting them to Eq. (\ref{Us}). As shown in Fig. \ref{1DHRPP_energetics}(a) (lower panel), there is good agreement between the DFT-calculated values and the fitted analytical relation within the elastic regime of $\pm$3\%; the root-mean-square error of the fit is less than 5 meV. We find that the edge elastic modulus $E_{\text{edge}}$ is 0.2651 eV/\AA$^2$, while the edge stress $f_{\text{edge}}$ is tensile and of magnitude 0.0108 eV/\AA$^2$. Although $E_{\text{edge}}$ and $f_{\text{edge}}$ have the same units, they are different physically. The edge elastic modulus reflects the work done to elastically deform an edge, whereas the edge stress refers to the work associated with varying the exposed surface at the edge.\cite{Cedstr94} Accordingly, a tensile (compressive) edge stress would mean that the edge has a tendency to undergo contraction (stretching). Edge stresses can therefore serve as an intrinsic driving mechanism for structural reconstructions and morphological changes at the layer--edge region. For example, compressive edge stresses have been identified as the origin of warping phenomena in graphene nanosheets.\cite{HLedgegr09,SRedgegr08}

In the present study, the edge stress of the (BA)$_2$PbI$_4$ hybrid perovskite is found to be tensile which induces contraction (or compression) at the edge. Since $f_{\text{edge}}$ is positive, for small enough compressive strain, the negative third term (linear to $\epsilon$) in Eq. (\ref{Us}) can always overcome the positive first and second terms ($\epsilon$ is of higher order) to make $U$\textsubscript{s} negative. So, the nanoribbon is unstable against a small amount of compression along the edge direction. Table \ref{relaxed_params} provides a summary of the results obtained across the $k = 1, 2, 3$ nanoribbons. As shown, the mixity factor can be as low as 0.6 for the narrowest of ribbons ($k$ = 1) and will tend to 1 beyond $k$ = 3. As the central 2D sheet-like region develops for wider nanoribbons, the critical values (minimum) of strain $\epsilon$\textsuperscript{cr} and strain energy $U$\textsubscript{s}\textsuperscript{cr} will decrease in magnitude accordingly due to edge pinning effects. Hence, the mechanical edge instability is expected to be strongly enhanced when thinning down the nanoribbon to its 1D limit.

The Young's modulus $E$ of monolayer (BA)$_2$PbI$_4$, as we calculate here, is 11--12 GPa. This is consistent with experimental results reported in a recent work by Tu \textit{et al.} where $E$ values in the range of 5--13 GPa were found for typical members of the (BA)$_2$(MA)$_{n-1}$Pb$_n$I$_{3n+1}$ family.\cite{TSultra2d18} Notably, compared to other common 2D materials, HRPPs are relatively soft; their Young's moduli are 1--2 orders of magnitude lower than those of graphene, h-BN, phosphorene, and MoS$_2$.\cite{FCgrbn17,WPphos14,BBmos211} This implies a greater flexibility of the hybrid perovskite nanostructure in response to external loading and reduced energy barriers of structural changes. Laboratory investigations of HRPPs have revealed the presence of highly unusual edge states exhibiting lower-energy photoluminescence (PL) emission.\cite{BTles17,QDhetedge20,ZTrevcon19,SDhumid19} PL imaging studies further showed that these states were not distributed continuously throughout the edges, suggesting that they were of non-native origin, and that the as-terminated edge structure was innately unstable. These edge instabilities can be rationalized in terms of the edge stress. For instance, tensile edge stresses effectively exert compressive external forces ($\tau_{\text{ext}}$) on the (BA)$_2$PbI$_4$ nanosheet, as illustrated in Figure \ref{1DHRPP_energetics}(b). These forces destabilize the layer--edge region, making it highly amenable to reconstructions into more energetically favorable states. In addition, the lattice mismatch between the edge and interior part of the nanosheet can promote the formation of voids and decouple the layer--edge structure at localized areas. Free edges, whether pre-existing or freshly created upon voiding, possess unsaturated metal/halide bonds and can present chemically active sites for the transformation into more stable configurations.

In principle, the edge stress can be tuned (and so, released) by modifying the chemical environment at the edges. This has been demonstrated in the case of graphene where hydrogen adsorption or Stone--Wales reconstructions at the edge can relieve the compressive edge stress and stabilize the planar structure.\cite{HLedgegr09} Reports on HRPPs with BA as the organic spacer have mainly alluded to two key processes that could possibly account for efficient edge stress dissipation: (i) BA-to-MA cation exchange reactions,\cite{QDhetedge20,ZTrevcon19} and (ii) water adsorption.\cite{SDhumid19} The specific passivation effects brought about by MA, H$_2$O, or even other small molecules on $f_{\text{edge}}$ are however beyond the scope of this study, but nonetheless, provide fertile ground for future work. One can also expect to regulate the edge stress with temperature. Heat treatments are known to produce structural reorientations and defects like organic vacancies, and could thus be an important step affecting the dynamics of edge stress relief mechanisms.\cite{LAplanar18}

\begin{table*}
\caption{Calculated width $w$, thickness $t$, mixity $\phi$, and critical values of strain $\epsilon$\textsuperscript{cr} and strain energy $U$\textsubscript{s}\textsuperscript{cr} of $k$ = 1, 2, 3 nanoribbons. \label{relaxed_params}}
\begin{ruledtabular}
{\renewcommand{\arraystretch}{1.5}
\begin{tabular}{cccccc}
System & $w$ (\AA) & $t$ (\AA) & $\phi$ & $\epsilon$\textsuperscript{cr} & $U$\textsubscript{s}\textsuperscript{cr} (meV) \\ \hline
$k$ = 1 & 13.06 & 15.57 & 0.61 & $-$1.98\% & $-$28.36 \\
$k$ = 2 & 21.85 & 15.59 & 0.89 & $-$1.11\% & $-$16.01 \\
$k$ = 3 & 30.64 & 15.59 & 0.99 & $-$0.79\% & $-$11.32
\end{tabular}}
\end{ruledtabular}
\end{table*}

\begin{figure*}
\includegraphics[width=0.5\textwidth]{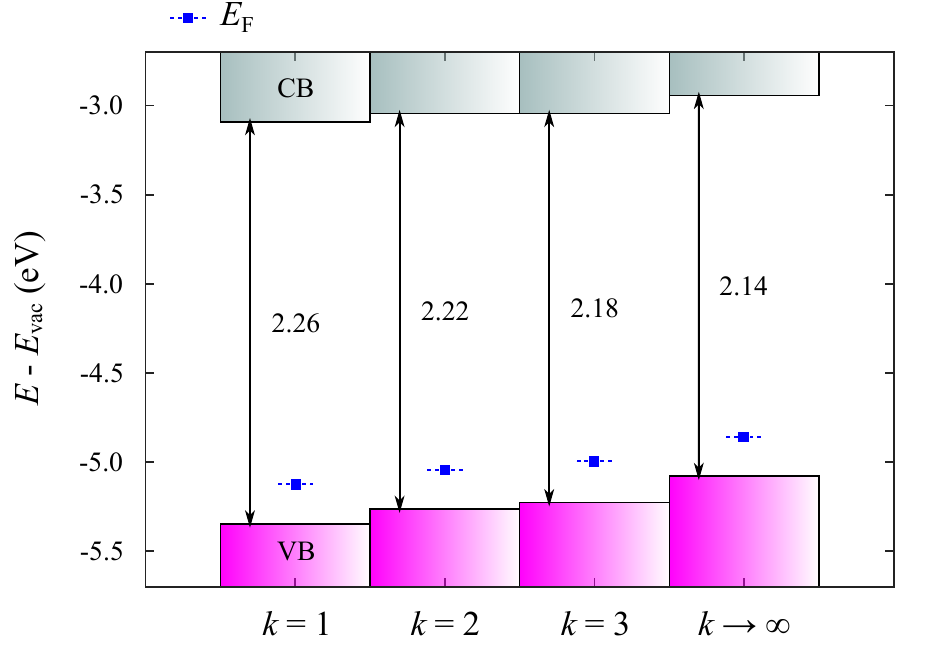}
\caption{Valence (pink) and conduction (grey) bands of $k$ = 1, 2, 3 nanoribbons and the 2D nanosheet ($k \rightarrow \infty$), aligned relative to the vacuum level $E_{\text{vac}}$. The Fermi level $E_{\text{F}}$ for each system is marked in blue. \label{HRPP_bands}}
\end{figure*}

The band alignment of the various nanoribbons as a function of width is also investigated, as shown in Fig. \ref{HRPP_bands}. For consistency, the 2D (BA)$_2$PbI$_4$ nanosheet is assigned a width index label of `$k \rightarrow \infty$', that is, an infinitely long nanoribbon. It is noted that the thinning of the nanoribbon from 2D to $k$ = 1 is accompanied by the widening of the band gap from 2.14 eV to 2.26 eV. All band gaps continue to retain their direct-gap character at the $\Gamma$ point (see Fig. S1 in the SI). Both the conduction band (CB) and valence band (VB), including the Fermi level $E_{\text{F}}$, are found to shift to higher energy levels as the width of the nanoribbon increases. This translates to increasing work functions of the hybrid perovskite as the width of the nanoribbon decreases. Compared to the equilibrium work function of 4.86 eV for the 2D (BA)$_2$PbI$_4$ nanosheet, the work function is calculated to be 5.13 eV for the $k$ = 1 nanoribbon. Our results show that the edges have lower aligned states compared to the pristine states of the 2D sheet. This implies that the edges can act as sinks for photo-excited electrons. In contrast, the holes tend to move towards the interior part of the layer, thus promoting electron--hole separation. These findings are in line with experimental measurements which have attested to the ability of edges to facilitate the dissociation of excitons into free carriers.\cite{BTles17,ZTrevcon19,SDhumid19}

\begin{figure*}
\includegraphics[width=1\textwidth]{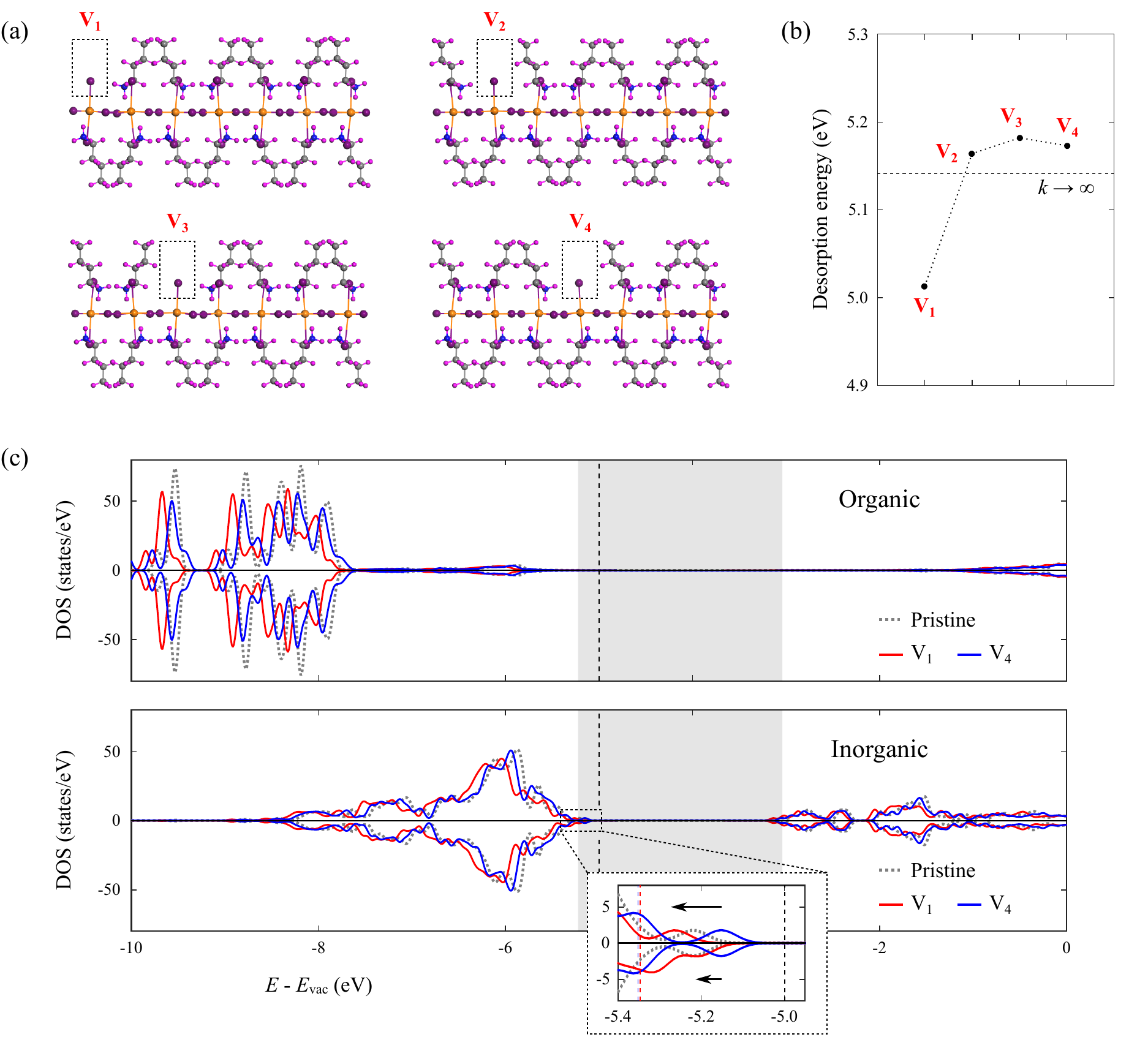}
\caption{(a) Relaxed structures of $k$ = 3 nanoribbons containing organic vacancies V$_i$ at different positions along its width, where $i$ = 1, 2, 3, 4. V$_1$ is located at the free edge while V$_4$ is situated in the central region of the nanoribbon. (b) Desorption energy of the organic molecule at the various vacancy sites. (c) DOS contributions from the organic and inorganic components of the nanoribbon containing no defects (pristine), an edge vacancy V$_1$, and a central vacancy V$_4$. The vacuum level $E_{\text{vac}}$ is set at zero and the Fermi levels are indicated by vertical dashed lines. The grey-shaded regions denote the band gap of the pristine nanoribbon. The inset provides a close-up view of the DOS peaks at the valence band edge. \label{HRPP_desorption}}
\end{figure*}

Finally, edge-induced instabilities can potentially alter the binding strength of the organic BA molecule to the inorganic [PbI$_6$]$^{4-}$ frame. In Fig. \ref{HRPP_desorption}(a), we consider the formation of organic vacancies along the $k$ = 3 nanoribbon. Note that this model is chosen because it allows for the simulation of defects at both the edge and central 2D sheet-like region. A vacancy at the free edge is denoted by V$_1$, while V$_2$, V$_3$ and V$_4$ are successive vacancy sites extending into the middle of the nanoribbon structure. Such vacancies typically form when the organic molecule is perturbed out of its range of hydrogen bonding and undergoes desorption from the perovskite. Figure \ref{HRPP_desorption}(b) shows the trend in desorption energy of the BA molecule at the different vacancy sites. The desorption energy is calculated as 5.14 eV in the 2D sheet (dashed line), which is comparable to that of V$_2$, V$_3$ and V$_4$ defects (5.16--5.18 eV) that lie in the inner part of the nanoribbon.

Interestingly, the desorption energy can be as low as 5.01 eV for V$_1$ at the free edge. This highlights significantly weakened interactions between the organic and inorganic components of the perovskite along edges and suggests that organic vacancies will tend to form at the edge rather than at the interior region of the nanosheet. Similar deductions have also been made in previous experiments, indicating that the loss of BA ligands do indeed readily occur at the edges.\cite{QDhetedge20,ZTrevcon19} The resulting vacancies can be vital for defect engineering since these sites can be decorated by alternative functional groups, that is of course, if they can bind well with the inorganic framework. BA-deficient edges may also be harnessed for gas sensing applications with the exposed [PbI$_6$]$^{4-}$ octahedra playing an active role by supporting charge transfer to/from the perovskite.

The density of states (DOS) of the pristine and defective nanoribbons are given in Fig. \ref{HRPP_desorption}(c). As shown, the regions near the Fermi level are dominated by states from the inorganic Pb and I atoms, whereas states from the organic molecule are highly deep lying. The localization of the organic (inorganic) states at regions far away from (near) the band edges is consistent with other types of hybrid perovskites,\cite{SCqwell16,WLdos18,SSdos17,BYdos14} signalling that the organic molecules do not directly affect the band gap. The DOS distribution we observe here for the pristine $k$ = 3 nanoribbon is verified to be consistent for all other nanoribbon widths and for the 2D nanosheet as well (see Fig. S2 in the SI). A closer inspection of the electronic states at the band edges reveal that the main contributions to the valence band maximum and conduction band minimum are from I-\textit{p} and Pb-\textit{p} orbitals, respectively (see Fig. S3 in the SI). The presence of vacancies is found to shift the energy level spectrum of monolayer (BA)$_2$PbI$_4$ towards the lower energy range. Evidently, an edge vacancy (V$_1$) causes a much more pronounced shift as compared to the central vacancy (V$_4$). A close-up view of the DOS at the valence band edge is provided in the inset of Fig. \ref{HRPP_desorption}(c). Although V$_4$ is a non-magnetic defect, V$_1$ is spin-polarized with a net magnetic moment of 0.48 $\mu$\textsubscript{B}. The downward shifts of the first DOS peak at the valence band edge between V$_1$ and V$_4$ are calculated to be 114 meV (spin-up) and 55 meV (spin-down). At the vacancy concentration investigated here ($\sim$7\%), both V$_1$ and V$_4$ are metallic. The intersection of the Fermi level with the upper part of the valence band suggests that organic vacancies can carry \textit{p}-type doping effects whereby defect states are introduced near the valence band edge. Moreover, the Fermi levels of the defective nanoribbons are lower, highlighting that these vacancies can lead to an increase in the work function, in this case, by $\sim$0.35 eV compared to the pristine structure.

\section{Conclusions}
In summary, we have contributed a first-principles study on the mechanical and electronic properties of edges of molecularly thin HRPPs by focusing on monolayer (BA)$_2$PbI$_4$ as a representative system. At edges, considerable structural distortions are apparent in the octahedra network, which could give rise to optoelectronic properties that are different from those of the 2D nanosheet. Mechanical characterization results indicate that the as-terminated edge structure is intrinsically unstable due to the existence of strong tensile edge stresses. In general, the edge stress may be released by modifying the chemical environment at the edge, for example through structural reorientations and/or defects such as vacancies and dopants. The edge stress can thus be responsible for driving reconstructions into more stable configurations. The precise relationship between the edge structure and edge stress is however not fully understood across hybrid perovskite compounds, and further research into the area will be a meaningful extension of the present work. Due to the rich compositional space of the edge structure, especially with the formation of defects and charged states, a high-volume screening approach will be needed for  efficient edge structure identification. This may be achieved by combining first-principles calculations with particle swarm optimization, machine learning, or cluster expansion techniques. In addition, it is shown that the edge can be associated with lower-energy electronic states that are conducive for the dissociation of photo-excited carriers, where electrons move towards the edge while holes move towards the interior part of the nanosheet. Our calculations also hint at the preferential desorption of organic BA molecules at the edge, in line with reports from previous experiments. The resulting edge vacancies could promote local substitutional doping of the halide/metal atom or serve as chemically active sites for the functionalization by other small molecules. It must also be pointed out that in Ruddlesden--Popper perovskites like (BA)$_2$PbI$_4$, unsaturated chemical bonds at the edge and organic vacancies do not create trap states inside the fundamental band gap, making them an appealing choice in the field of photonics. Overall, the findings reported here shed light on the underlying mechanisms responsible for edge states in HRPPs and cater to both theoretical and experimental interests to develop high-performance layer--edge devices.

\section{Methods}
Spin-polarized first-principles calculations are performed within the framework of DFT \cite{KSexcorr65,SSdft11} using the Vienna \textit{ab initio} simulation package (VASP).\cite{KFvasp96} The Perdew--Burke--Ernzerhof (PBE) \cite{PBE96} exchange--correlation functional is adopted in this work along with the DFT-D2 method of Grimme \cite{DFTD206} to account for the van der Waals forces in 2D-layered (BA)$_2$PbI$_4$. A kinetic energy cutoff of 500 eV is selected for the plane wave basis set. Both nanosheets (2D) and nanoribbons (1D) are modelled in this study by supplying suitable boundary conditions to the simulation cell. In the case of 2D nanosheets, periodic boundary conditions (PBCs) are applied in the in-plane (\textit{x}, \textit{y}) directions, whereas free boundary conditions (FBCs) are imposed in the normal (\textit{z}) direction by introducing a vacuum separation layer of $\sim$15 \AA\ between neighboring slabs. Correspondingly, 1D nanoribbons are simulated by applying PBCs only along the length direction, whilst FBCs are enforced by adding vacuum regions of $\sim$15 \AA\ along both the normal and width directions. The investigated nanosheets and edge A (B) nanoribbons are sampled in the Brillouin zone with 4 $\times$ 4 $\times$ 1 and 4 $\times$ 1 $\times$ 1 (1 $\times$ 4 $\times$ 1) \textit{k}-point grids, respectively, using the Monkhorst--Pack method. The energy convergence criteria for electronic iterations is set at 10$^{-6}$ eV, and all structures are relaxed until the maximum Hellmann--Feynman force per atom is smaller than 0.02 eV/\AA. By convention, strained conditions are prescribed according to the relation \textit{a}$_1$ = \textit{a}$_0$(1+$\epsilon$), where \textit{a}$_0$ and \textit{a}$_1$ are the lattice constants in equilibrium and at a strain level of $\epsilon$, respectively.

\section{Associated content}
\subsection{Supporting Information}
The Supporting Information is available free of charge at [\textit{insert link here}]. The file contains the formulation adopted for elastic constants characterization and DFT simulation results (electronic band structures, density of states, and electron density contour plots) accessory to the main article.

\subsection{Notes}
The authors declare no competing financial interest.

\section{Acknowledgments}
The authors acknowledge the financial support received from the Nanyang Environment and Water Research Institute (Core Funding), Nanyang Technological University, Singapore. The computational work for this article was partially performed on resources of the National Supercomputing Centre, Singapore (https://www.nscc.sg). Y. Cai acknowledges the support provided by the University of Macau (SRG2019-00179-IAPME), the Science and Technology Development Fund from Macau SAR (FDCT-0163/2019/A3), the Natural Science Foundation of China (Grant 22022309) and the Natural Science Foundation of Guangdong Province, China (2021A1515010024).

\bibliography{bib_manuscript}

\providecommand{\noopsort}[1]{}\providecommand{\singleletter}[1]{#1}%
\begin{thebibliography}{57}%
\makeatletter
\providecommand \@ifxundefined [1]{%
 \@ifx{#1\undefined}
}%
\providecommand \@ifnum [1]{%
 \ifnum #1\expandafter \@firstoftwo
 \else \expandafter \@secondoftwo
 \fi
}%
\providecommand \@ifx [1]{%
 \ifx #1\expandafter \@firstoftwo
 \else \expandafter \@secondoftwo
 \fi
}%
\providecommand \natexlab [1]{#1}%
\providecommand \enquote  [1]{``#1''}%
\providecommand \bibnamefont  [1]{#1}%
\providecommand \bibfnamefont [1]{#1}%
\providecommand \citenamefont [1]{#1}%
\providecommand \href@noop [0]{\@secondoftwo}%
\providecommand \href [0]{\begingroup \@sanitize@url \@href}%
\providecommand \@href[1]{\@@startlink{#1}\@@href}%
\providecommand \@@href[1]{\endgroup#1\@@endlink}%
\providecommand \@sanitize@url [0]{\catcode `\\12\catcode `\$12\catcode
  `\&12\catcode `\#12\catcode `\^12\catcode `\_12\catcode `\%12\relax}%
\providecommand \@@startlink[1]{}%
\providecommand \@@endlink[0]{}%
\providecommand \url  [0]{\begingroup\@sanitize@url \@url }%
\providecommand \@url [1]{\endgroup\@href {#1}{\urlprefix }}%
\providecommand \urlprefix  [0]{URL }%
\providecommand \Eprint [0]{\href }%
\providecommand \doibase [0]{https://doi.org/}%
\providecommand \selectlanguage [0]{\@gobble}%
\providecommand \bibinfo  [0]{\@secondoftwo}%
\providecommand \bibfield  [0]{\@secondoftwo}%
\providecommand \translation [1]{[#1]}%
\providecommand \BibitemOpen [0]{}%
\providecommand \bibitemStop [0]{}%
\providecommand \bibitemNoStop [0]{.\EOS\space}%
\providecommand \EOS [0]{\spacefactor3000\relax}%
\providecommand \BibitemShut  [1]{\csname bibitem#1\endcsname}%
\let\auto@bib@innerbib\@empty
\bibitem [{\citenamefont {Matsushima}\ \emph {et~al.}(2016)\citenamefont
  {Matsushima}, \citenamefont {Hwang}, \citenamefont {Sandanayaka},
  \citenamefont {Qin}, \citenamefont {Terakawa}, \citenamefont {Fujihara},
  \citenamefont {Yahiro},\ and\ \citenamefont {Adachi}}]{MHhrpp16}%
  \BibitemOpen
  \bibfield  {author} {\bibinfo {author} {\bibfnamefont {T.}~\bibnamefont
  {Matsushima}}, \bibinfo {author} {\bibfnamefont {S.}~\bibnamefont {Hwang}},
  \bibinfo {author} {\bibfnamefont {A.~S.~D.}\ \bibnamefont {Sandanayaka}},
  \bibinfo {author} {\bibfnamefont {C.}~\bibnamefont {Qin}}, \bibinfo {author}
  {\bibfnamefont {S.}~\bibnamefont {Terakawa}}, \bibinfo {author}
  {\bibfnamefont {T.}~\bibnamefont {Fujihara}}, \bibinfo {author}
  {\bibfnamefont {M.}~\bibnamefont {Yahiro}},\ and\ \bibinfo {author}
  {\bibfnamefont {C.}~\bibnamefont {Adachi}},\ }\bibfield  {title} {\bibinfo
  {title} {{Solution-Processed Organic-Inorganic Perovskite Field-Effect
  Transistors with High Hole Mobilities}},\ }\href
  {https://doi.org/https://doi.org/10.1002/adma.201603126} {\bibfield
  {journal} {\bibinfo  {journal} {Adv. Mater.}\ }\textbf {\bibinfo {volume}
  {28}},\ \bibinfo {pages} {10275} (\bibinfo {year} {2016})}\BibitemShut
  {NoStop}%
\bibitem [{\citenamefont {Tsai}\ \emph {et~al.}(2016)\citenamefont {Tsai},
  \citenamefont {Nie}, \citenamefont {Blancon}, \citenamefont {Stoumpos},
  \citenamefont {Asadpour}, \citenamefont {Harutyunyan}, \citenamefont
  {Neukirch}, \citenamefont {Verduzco}, \citenamefont {Crochet}, \citenamefont
  {Tretiak}, \citenamefont {Pedesseau}, \citenamefont {Even}, \citenamefont
  {Alam}, \citenamefont {Gupta}, \citenamefont {Lou}, \citenamefont {Ajayan},
  \citenamefont {Bedzyk}, \citenamefont {Kanatzidis},\ and\ \citenamefont
  {Mohite}}]{TNhrpp16}%
  \BibitemOpen
  \bibfield  {author} {\bibinfo {author} {\bibfnamefont {H.}~\bibnamefont
  {Tsai}}, \bibinfo {author} {\bibfnamefont {W.}~\bibnamefont {Nie}}, \bibinfo
  {author} {\bibfnamefont {J.-C.}\ \bibnamefont {Blancon}}, \bibinfo {author}
  {\bibfnamefont {C.~C.}\ \bibnamefont {Stoumpos}}, \bibinfo {author}
  {\bibfnamefont {R.}~\bibnamefont {Asadpour}}, \bibinfo {author}
  {\bibfnamefont {B.}~\bibnamefont {Harutyunyan}}, \bibinfo {author}
  {\bibfnamefont {A.~J.}\ \bibnamefont {Neukirch}}, \bibinfo {author}
  {\bibfnamefont {R.}~\bibnamefont {Verduzco}}, \bibinfo {author}
  {\bibfnamefont {J.~J.}\ \bibnamefont {Crochet}}, \bibinfo {author}
  {\bibfnamefont {S.}~\bibnamefont {Tretiak}}, \bibinfo {author} {\bibfnamefont
  {L.}~\bibnamefont {Pedesseau}}, \bibinfo {author} {\bibfnamefont
  {J.}~\bibnamefont {Even}}, \bibinfo {author} {\bibfnamefont {M.~A.}\
  \bibnamefont {Alam}}, \bibinfo {author} {\bibfnamefont {G.}~\bibnamefont
  {Gupta}}, \bibinfo {author} {\bibfnamefont {J.}~\bibnamefont {Lou}}, \bibinfo
  {author} {\bibfnamefont {P.~M.}\ \bibnamefont {Ajayan}}, \bibinfo {author}
  {\bibfnamefont {M.~J.}\ \bibnamefont {Bedzyk}}, \bibinfo {author}
  {\bibfnamefont {M.~G.}\ \bibnamefont {Kanatzidis}},\ and\ \bibinfo {author}
  {\bibfnamefont {A.~D.}\ \bibnamefont {Mohite}},\ }\bibfield  {title}
  {\bibinfo {title} {{High-Efficiency Two-Dimensional Ruddlesden-Popper
  Perovskite Solar Cells}},\ }\href {https://doi.org/10.1038/nature18306}
  {\bibfield  {journal} {\bibinfo  {journal} {Nature}\ }\textbf {\bibinfo
  {volume} {536}},\ \bibinfo {pages} {312} (\bibinfo {year}
  {2016})}\BibitemShut {NoStop}%
\bibitem [{\citenamefont {Cao}\ \emph {et~al.}(2015{\natexlab{a}})\citenamefont
  {Cao}, \citenamefont {Stoumpos}, \citenamefont {Farha}, \citenamefont
  {Hupp},\ and\ \citenamefont {Kanatzidis}}]{CShrpp15}%
  \BibitemOpen
  \bibfield  {author} {\bibinfo {author} {\bibfnamefont {D.~H.}\ \bibnamefont
  {Cao}}, \bibinfo {author} {\bibfnamefont {C.~C.}\ \bibnamefont {Stoumpos}},
  \bibinfo {author} {\bibfnamefont {O.~K.}\ \bibnamefont {Farha}}, \bibinfo
  {author} {\bibfnamefont {J.~T.}\ \bibnamefont {Hupp}},\ and\ \bibinfo
  {author} {\bibfnamefont {M.~G.}\ \bibnamefont {Kanatzidis}},\ }\bibfield
  {title} {\bibinfo {title} {{2D Homologous Perovskites as Light-Absorbing
  Materials for Solar Cell Applications}},\ }\href
  {https://doi.org/10.1021/jacs.5b03796} {\bibfield  {journal} {\bibinfo
  {journal} {J. Am. Chem. Soc.}\ }\textbf {\bibinfo {volume} {137}},\ \bibinfo
  {pages} {7843} (\bibinfo {year} {2015}{\natexlab{a}})}\BibitemShut {NoStop}%
\bibitem [{\citenamefont {Smith}\ \emph {et~al.}(2014)\citenamefont {Smith},
  \citenamefont {Hoke}, \citenamefont {Solis-Ibarra}, \citenamefont {McGehee},\
  and\ \citenamefont {Karunadasa}}]{SHhrpp14}%
  \BibitemOpen
  \bibfield  {author} {\bibinfo {author} {\bibfnamefont {I.~C.}\ \bibnamefont
  {Smith}}, \bibinfo {author} {\bibfnamefont {E.~T.}\ \bibnamefont {Hoke}},
  \bibinfo {author} {\bibfnamefont {D.}~\bibnamefont {Solis-Ibarra}}, \bibinfo
  {author} {\bibfnamefont {M.~D.}\ \bibnamefont {McGehee}},\ and\ \bibinfo
  {author} {\bibfnamefont {H.~I.}\ \bibnamefont {Karunadasa}},\ }\bibfield
  {title} {\bibinfo {title} {{A Layered Hybrid Perovskite Solar-Cell Absorber
  with Enhanced Moisture Stability}},\ }\href
  {https://doi.org/https://doi.org/10.1002/anie.201406466} {\bibfield
  {journal} {\bibinfo  {journal} {Angew. Chem. Int. Ed.}\ }\textbf {\bibinfo
  {volume} {53}},\ \bibinfo {pages} {11232} (\bibinfo {year}
  {2014})}\BibitemShut {NoStop}%
\bibitem [{\citenamefont {Kagan}\ \emph {et~al.}(1999)\citenamefont {Kagan},
  \citenamefont {Mitzi},\ and\ \citenamefont {Dimitrakopoulos}}]{KMhrpp99}%
  \BibitemOpen
  \bibfield  {author} {\bibinfo {author} {\bibfnamefont {C.~R.}\ \bibnamefont
  {Kagan}}, \bibinfo {author} {\bibfnamefont {D.~B.}\ \bibnamefont {Mitzi}},\
  and\ \bibinfo {author} {\bibfnamefont {C.~D.}\ \bibnamefont
  {Dimitrakopoulos}},\ }\bibfield  {title} {\bibinfo {title}
  {{Organic-Inorganic Hybrid Materials as Semiconducting Channels in Thin-Film
  Field-Effect Transistors}},\ }\href
  {https://doi.org/10.1126/science.286.5441.945} {\bibfield  {journal}
  {\bibinfo  {journal} {Science}\ }\textbf {\bibinfo {volume} {286}},\ \bibinfo
  {pages} {945} (\bibinfo {year} {1999})}\BibitemShut {NoStop}%
\bibitem [{\citenamefont {Muljarov}\ \emph {et~al.}(1995)\citenamefont
  {Muljarov}, \citenamefont {Tikhodeev}, \citenamefont {Gippius},\ and\
  \citenamefont {Ishihara}}]{MTmqw95}%
  \BibitemOpen
  \bibfield  {author} {\bibinfo {author} {\bibfnamefont {E.~A.}\ \bibnamefont
  {Muljarov}}, \bibinfo {author} {\bibfnamefont {S.~G.}\ \bibnamefont
  {Tikhodeev}}, \bibinfo {author} {\bibfnamefont {N.~A.}\ \bibnamefont
  {Gippius}},\ and\ \bibinfo {author} {\bibfnamefont {T.}~\bibnamefont
  {Ishihara}},\ }\bibfield  {title} {\bibinfo {title} {{Excitons in
  Self-Organized Semiconductor/Insulator Superlattices: PbI-Based Perovskite
  Compounds}},\ }\href {https://doi.org/10.1103/PhysRevB.51.14370} {\bibfield
  {journal} {\bibinfo  {journal} {Phys. Rev. B}\ }\textbf {\bibinfo {volume}
  {51}},\ \bibinfo {pages} {14370} (\bibinfo {year} {1995})}\BibitemShut
  {NoStop}%
\bibitem [{\citenamefont {Hong}\ \emph {et~al.}(1992)\citenamefont {Hong},
  \citenamefont {Ishihara},\ and\ \citenamefont {Nurmikko}}]{HImqw92}%
  \BibitemOpen
  \bibfield  {author} {\bibinfo {author} {\bibfnamefont {X.}~\bibnamefont
  {Hong}}, \bibinfo {author} {\bibfnamefont {T.}~\bibnamefont {Ishihara}},\
  and\ \bibinfo {author} {\bibfnamefont {A.~V.}\ \bibnamefont {Nurmikko}},\
  }\bibfield  {title} {\bibinfo {title} {{Dielectric Confinement Effect on
  Excitons in ${\mathrm{PbI}}_{4}$-Based Layered Semiconductors}},\ }\href
  {https://doi.org/10.1103/PhysRevB.45.6961} {\bibfield  {journal} {\bibinfo
  {journal} {Phys. Rev. B}\ }\textbf {\bibinfo {volume} {45}},\ \bibinfo
  {pages} {6961} (\bibinfo {year} {1992})}\BibitemShut {NoStop}%
\bibitem [{\citenamefont {Blancon}\ \emph {et~al.}(2018)\citenamefont
  {Blancon}, \citenamefont {Stier}, \citenamefont {Tsai}, \citenamefont {Nie},
  \citenamefont {Stoumpos}, \citenamefont {Traor{\'e}}, \citenamefont
  {Pedesseau}, \citenamefont {Kepenekian}, \citenamefont {Katsutani},
  \citenamefont {Noe}, \citenamefont {Kono}, \citenamefont {Tretiak},
  \citenamefont {Crooker}, \citenamefont {Katan}, \citenamefont {Kanatzidis},
  \citenamefont {Crochet}, \citenamefont {Even},\ and\ \citenamefont
  {Mohite}}]{BAqwell18}%
  \BibitemOpen
  \bibfield  {author} {\bibinfo {author} {\bibfnamefont {J.-C.}\ \bibnamefont
  {Blancon}}, \bibinfo {author} {\bibfnamefont {A.~V.}\ \bibnamefont {Stier}},
  \bibinfo {author} {\bibfnamefont {H.}~\bibnamefont {Tsai}}, \bibinfo {author}
  {\bibfnamefont {W.}~\bibnamefont {Nie}}, \bibinfo {author} {\bibfnamefont
  {C.~C.}\ \bibnamefont {Stoumpos}}, \bibinfo {author} {\bibfnamefont
  {B.}~\bibnamefont {Traor{\'e}}}, \bibinfo {author} {\bibfnamefont
  {L.}~\bibnamefont {Pedesseau}}, \bibinfo {author} {\bibfnamefont
  {M.}~\bibnamefont {Kepenekian}}, \bibinfo {author} {\bibfnamefont
  {F.}~\bibnamefont {Katsutani}}, \bibinfo {author} {\bibfnamefont {G.~T.}\
  \bibnamefont {Noe}}, \bibinfo {author} {\bibfnamefont {J.}~\bibnamefont
  {Kono}}, \bibinfo {author} {\bibfnamefont {S.}~\bibnamefont {Tretiak}},
  \bibinfo {author} {\bibfnamefont {S.~A.}\ \bibnamefont {Crooker}}, \bibinfo
  {author} {\bibfnamefont {C.}~\bibnamefont {Katan}}, \bibinfo {author}
  {\bibfnamefont {M.~G.}\ \bibnamefont {Kanatzidis}}, \bibinfo {author}
  {\bibfnamefont {J.~J.}\ \bibnamefont {Crochet}}, \bibinfo {author}
  {\bibfnamefont {J.}~\bibnamefont {Even}},\ and\ \bibinfo {author}
  {\bibfnamefont {A.~D.}\ \bibnamefont {Mohite}},\ }\bibfield  {title}
  {\bibinfo {title} {{Scaling Law for Excitons in 2D Perovskite Quantum
  Wells}},\ }\href {https://doi.org/10.1038/s41467-018-04659-x} {\bibfield
  {journal} {\bibinfo  {journal} {Nat. Commun.}\ }\textbf {\bibinfo {volume}
  {9}},\ \bibinfo {pages} {2254} (\bibinfo {year} {2018})}\BibitemShut
  {NoStop}%
\bibitem [{\citenamefont {Gao}\ \emph {et~al.}(2018)\citenamefont {Gao},
  \citenamefont {Bin Mohd~Yusoff},\ and\ \citenamefont
  {Nazeeruddin}}]{GBqwell18}%
  \BibitemOpen
  \bibfield  {author} {\bibinfo {author} {\bibfnamefont {P.}~\bibnamefont
  {Gao}}, \bibinfo {author} {\bibfnamefont {A.~R.}\ \bibnamefont {Bin
  Mohd~Yusoff}},\ and\ \bibinfo {author} {\bibfnamefont {M.~K.}\ \bibnamefont
  {Nazeeruddin}},\ }\bibfield  {title} {\bibinfo {title} {{Dimensionality
  Engineering of Hybrid Halide Perovskite Light Absorbers}},\ }\href
  {https://doi.org/10.1038/s41467-018-07382-9} {\bibfield  {journal} {\bibinfo
  {journal} {Nat. Commun.}\ }\textbf {\bibinfo {volume} {9}},\ \bibinfo {pages}
  {5028} (\bibinfo {year} {2018})}\BibitemShut {NoStop}%
\bibitem [{\citenamefont {Xiao}\ \emph {et~al.}(2017)\citenamefont {Xiao},
  \citenamefont {Meng}, \citenamefont {Wang}, \citenamefont {Mitzi},\ and\
  \citenamefont {Yan}}]{XMqwell17}%
  \BibitemOpen
  \bibfield  {author} {\bibinfo {author} {\bibfnamefont {Z.}~\bibnamefont
  {Xiao}}, \bibinfo {author} {\bibfnamefont {W.}~\bibnamefont {Meng}}, \bibinfo
  {author} {\bibfnamefont {J.}~\bibnamefont {Wang}}, \bibinfo {author}
  {\bibfnamefont {D.~B.}\ \bibnamefont {Mitzi}},\ and\ \bibinfo {author}
  {\bibfnamefont {Y.}~\bibnamefont {Yan}},\ }\bibfield  {title} {\bibinfo
  {title} {{Searching for Promising New Perovskite-Based Photovoltaic
  Absorbers: The Importance of Electronic Dimensionality}},\ }\href
  {https://doi.org/10.1039/C6MH00519E} {\bibfield  {journal} {\bibinfo
  {journal} {Mater. Horiz.}\ }\textbf {\bibinfo {volume} {4}},\ \bibinfo
  {pages} {206} (\bibinfo {year} {2017})}\BibitemShut {NoStop}%
\bibitem [{\citenamefont {Quan}\ \emph {et~al.}(2016)\citenamefont {Quan},
  \citenamefont {Yuan}, \citenamefont {Comin}, \citenamefont {Voznyy},
  \citenamefont {Beauregard}, \citenamefont {Hoogland}, \citenamefont {Buin},
  \citenamefont {Kirmani}, \citenamefont {Zhao}, \citenamefont {Amassian},
  \citenamefont {Kim},\ and\ \citenamefont {Sargent}}]{QYqwell16}%
  \BibitemOpen
  \bibfield  {author} {\bibinfo {author} {\bibfnamefont {L.~N.}\ \bibnamefont
  {Quan}}, \bibinfo {author} {\bibfnamefont {M.}~\bibnamefont {Yuan}}, \bibinfo
  {author} {\bibfnamefont {R.}~\bibnamefont {Comin}}, \bibinfo {author}
  {\bibfnamefont {O.}~\bibnamefont {Voznyy}}, \bibinfo {author} {\bibfnamefont
  {E.~M.}\ \bibnamefont {Beauregard}}, \bibinfo {author} {\bibfnamefont
  {S.}~\bibnamefont {Hoogland}}, \bibinfo {author} {\bibfnamefont
  {A.}~\bibnamefont {Buin}}, \bibinfo {author} {\bibfnamefont {A.~R.}\
  \bibnamefont {Kirmani}}, \bibinfo {author} {\bibfnamefont {K.}~\bibnamefont
  {Zhao}}, \bibinfo {author} {\bibfnamefont {A.}~\bibnamefont {Amassian}},
  \bibinfo {author} {\bibfnamefont {D.~H.}\ \bibnamefont {Kim}},\ and\ \bibinfo
  {author} {\bibfnamefont {E.~H.}\ \bibnamefont {Sargent}},\ }\bibfield
  {title} {\bibinfo {title} {{Ligand-Stabilized Reduced-Dimensionality
  Perovskites}},\ }\href {https://doi.org/10.1021/jacs.5b11740} {\bibfield
  {journal} {\bibinfo  {journal} {J. Am. Chem. Soc.}\ }\textbf {\bibinfo
  {volume} {138}},\ \bibinfo {pages} {2649} (\bibinfo {year}
  {2016})}\BibitemShut {NoStop}%
\bibitem [{\citenamefont {Stoumpos}\ \emph {et~al.}(2016)\citenamefont
  {Stoumpos}, \citenamefont {Cao}, \citenamefont {Clark}, \citenamefont
  {Young}, \citenamefont {Rondinelli}, \citenamefont {Jang}, \citenamefont
  {Hupp},\ and\ \citenamefont {Kanatzidis}}]{SCqwell16}%
  \BibitemOpen
  \bibfield  {author} {\bibinfo {author} {\bibfnamefont {C.~C.}\ \bibnamefont
  {Stoumpos}}, \bibinfo {author} {\bibfnamefont {D.~H.}\ \bibnamefont {Cao}},
  \bibinfo {author} {\bibfnamefont {D.~J.}\ \bibnamefont {Clark}}, \bibinfo
  {author} {\bibfnamefont {J.}~\bibnamefont {Young}}, \bibinfo {author}
  {\bibfnamefont {J.~M.}\ \bibnamefont {Rondinelli}}, \bibinfo {author}
  {\bibfnamefont {J.~I.}\ \bibnamefont {Jang}}, \bibinfo {author}
  {\bibfnamefont {J.~T.}\ \bibnamefont {Hupp}},\ and\ \bibinfo {author}
  {\bibfnamefont {M.~G.}\ \bibnamefont {Kanatzidis}},\ }\bibfield  {title}
  {\bibinfo {title} {{Ruddlesden-Popper Hybrid Lead Iodide Perovskite 2D
  Homologous Semiconductors}},\ }\href
  {https://doi.org/10.1021/acs.chemmater.6b00847} {\bibfield  {journal}
  {\bibinfo  {journal} {Chem. Mater.}\ }\textbf {\bibinfo {volume} {28}},\
  \bibinfo {pages} {2852} (\bibinfo {year} {2016})}\BibitemShut {NoStop}%
\bibitem [{\citenamefont {Spanopoulos}\ \emph {et~al.}(2019)\citenamefont
  {Spanopoulos}, \citenamefont {Hadar}, \citenamefont {Ke}, \citenamefont {Tu},
  \citenamefont {Chen}, \citenamefont {Tsai}, \citenamefont {He}, \citenamefont
  {Shekhawat}, \citenamefont {Dravid}, \citenamefont {Wasielewski},
  \citenamefont {Mohite}, \citenamefont {Stoumpos},\ and\ \citenamefont
  {Kanatzidis}}]{SHstab19}%
  \BibitemOpen
  \bibfield  {author} {\bibinfo {author} {\bibfnamefont {I.}~\bibnamefont
  {Spanopoulos}}, \bibinfo {author} {\bibfnamefont {I.}~\bibnamefont {Hadar}},
  \bibinfo {author} {\bibfnamefont {W.}~\bibnamefont {Ke}}, \bibinfo {author}
  {\bibfnamefont {Q.}~\bibnamefont {Tu}}, \bibinfo {author} {\bibfnamefont
  {M.}~\bibnamefont {Chen}}, \bibinfo {author} {\bibfnamefont {H.}~\bibnamefont
  {Tsai}}, \bibinfo {author} {\bibfnamefont {Y.}~\bibnamefont {He}}, \bibinfo
  {author} {\bibfnamefont {G.}~\bibnamefont {Shekhawat}}, \bibinfo {author}
  {\bibfnamefont {V.~P.}\ \bibnamefont {Dravid}}, \bibinfo {author}
  {\bibfnamefont {M.~R.}\ \bibnamefont {Wasielewski}}, \bibinfo {author}
  {\bibfnamefont {A.~D.}\ \bibnamefont {Mohite}}, \bibinfo {author}
  {\bibfnamefont {C.~C.}\ \bibnamefont {Stoumpos}},\ and\ \bibinfo {author}
  {\bibfnamefont {M.~G.}\ \bibnamefont {Kanatzidis}},\ }\bibfield  {title}
  {\bibinfo {title} {{Uniaxial Expansion of the 2D Ruddlesden-Popper Perovskite
  Family for Improved Environmental Stability}},\ }\href
  {https://doi.org/10.1021/jacs.9b01327} {\bibfield  {journal} {\bibinfo
  {journal} {J. Am. Chem. Soc.}\ }\textbf {\bibinfo {volume} {141}},\ \bibinfo
  {pages} {5518} (\bibinfo {year} {2019})}\BibitemShut {NoStop}%
\bibitem [{\citenamefont {Onoda-Yamamuro}\ \emph {et~al.}(1990)\citenamefont
  {Onoda-Yamamuro}, \citenamefont {Matsuo},\ and\ \citenamefont
  {Suga}}]{OM3dphase90}%
  \BibitemOpen
  \bibfield  {author} {\bibinfo {author} {\bibfnamefont {N.}~\bibnamefont
  {Onoda-Yamamuro}}, \bibinfo {author} {\bibfnamefont {T.}~\bibnamefont
  {Matsuo}},\ and\ \bibinfo {author} {\bibfnamefont {H.}~\bibnamefont {Suga}},\
  }\bibfield  {title} {\bibinfo {title} {{Calorimetric and IR Spectroscopic
  Studies of Phase Transitions in Methylammonium Trihalogenoplumbates (II)}},\
  }\href {https://doi.org/https://doi.org/10.1016/0022-3697(90)90021-7}
  {\bibfield  {journal} {\bibinfo  {journal} {J. Phys. Chem. Solids}\ }\textbf
  {\bibinfo {volume} {51}},\ \bibinfo {pages} {1383} (\bibinfo {year}
  {1990})}\BibitemShut {NoStop}%
\bibitem [{\citenamefont {Poglitsch}\ and\ \citenamefont
  {Weber}(1987)}]{PW3dphase87}%
  \BibitemOpen
  \bibfield  {author} {\bibinfo {author} {\bibfnamefont {A.}~\bibnamefont
  {Poglitsch}}\ and\ \bibinfo {author} {\bibfnamefont {D.}~\bibnamefont
  {Weber}},\ }\bibfield  {title} {\bibinfo {title} {{Dynamic Disorder in
  Methylammoniumtrihalogenoplumbates (II) Observed by Millimeter-Wave
  Spectroscopy}},\ }\href {https://doi.org/10.1063/1.453467} {\bibfield
  {journal} {\bibinfo  {journal} {J. Chem. Phys.}\ }\textbf {\bibinfo {volume}
  {87}},\ \bibinfo {pages} {6373} (\bibinfo {year} {1987})}\BibitemShut
  {NoStop}%
\bibitem [{\citenamefont {Fang}\ \emph {et~al.}(2019)\citenamefont {Fang},
  \citenamefont {Wang}, \citenamefont {Shen}, \citenamefont {Shen},
  \citenamefont {Wang}, \citenamefont {Ma}, \citenamefont {Wang}, \citenamefont
  {Luo},\ and\ \citenamefont {Li}}]{FWplanar19}%
  \BibitemOpen
  \bibfield  {author} {\bibinfo {author} {\bibfnamefont {C.}~\bibnamefont
  {Fang}}, \bibinfo {author} {\bibfnamefont {H.}~\bibnamefont {Wang}}, \bibinfo
  {author} {\bibfnamefont {Z.}~\bibnamefont {Shen}}, \bibinfo {author}
  {\bibfnamefont {H.}~\bibnamefont {Shen}}, \bibinfo {author} {\bibfnamefont
  {S.}~\bibnamefont {Wang}}, \bibinfo {author} {\bibfnamefont {J.}~\bibnamefont
  {Ma}}, \bibinfo {author} {\bibfnamefont {J.}~\bibnamefont {Wang}}, \bibinfo
  {author} {\bibfnamefont {H.}~\bibnamefont {Luo}},\ and\ \bibinfo {author}
  {\bibfnamefont {D.}~\bibnamefont {Li}},\ }\bibfield  {title} {\bibinfo
  {title} {{High-Performance Photodetectors Based on Lead-Free 2D
  Ruddlesden-Popper Perovskite/MoS$_2$ Heterostructures}},\ }\href
  {https://doi.org/10.1021/acsami.8b20538} {\bibfield  {journal} {\bibinfo
  {journal} {ACS Appl. Mater. Interfaces}\ }\textbf {\bibinfo {volume} {11}},\
  \bibinfo {pages} {8419} (\bibinfo {year} {2019})}\BibitemShut {NoStop}%
\bibitem [{\citenamefont {Leng}\ \emph {et~al.}(2018)\citenamefont {Leng},
  \citenamefont {Abdelwahab}, \citenamefont {Verzhbitskiy}, \citenamefont
  {Telychko}, \citenamefont {Chu}, \citenamefont {Fu}, \citenamefont {Chi},
  \citenamefont {Guo}, \citenamefont {Chen}, \citenamefont {Chen},
  \citenamefont {Zhang}, \citenamefont {Xu}, \citenamefont {Lu}, \citenamefont
  {Chhowalla}, \citenamefont {Eda},\ and\ \citenamefont {Loh}}]{LAplanar18}%
  \BibitemOpen
  \bibfield  {author} {\bibinfo {author} {\bibfnamefont {K.}~\bibnamefont
  {Leng}}, \bibinfo {author} {\bibfnamefont {I.}~\bibnamefont {Abdelwahab}},
  \bibinfo {author} {\bibfnamefont {I.}~\bibnamefont {Verzhbitskiy}}, \bibinfo
  {author} {\bibfnamefont {M.}~\bibnamefont {Telychko}}, \bibinfo {author}
  {\bibfnamefont {L.}~\bibnamefont {Chu}}, \bibinfo {author} {\bibfnamefont
  {W.}~\bibnamefont {Fu}}, \bibinfo {author} {\bibfnamefont {X.}~\bibnamefont
  {Chi}}, \bibinfo {author} {\bibfnamefont {N.}~\bibnamefont {Guo}}, \bibinfo
  {author} {\bibfnamefont {Z.}~\bibnamefont {Chen}}, \bibinfo {author}
  {\bibfnamefont {Z.}~\bibnamefont {Chen}}, \bibinfo {author} {\bibfnamefont
  {C.}~\bibnamefont {Zhang}}, \bibinfo {author} {\bibfnamefont {Q.-H.}\
  \bibnamefont {Xu}}, \bibinfo {author} {\bibfnamefont {J.}~\bibnamefont {Lu}},
  \bibinfo {author} {\bibfnamefont {M.}~\bibnamefont {Chhowalla}}, \bibinfo
  {author} {\bibfnamefont {G.}~\bibnamefont {Eda}},\ and\ \bibinfo {author}
  {\bibfnamefont {K.~P.}\ \bibnamefont {Loh}},\ }\bibfield  {title} {\bibinfo
  {title} {{Molecularly Thin Two-Dimensional Hybrid Perovskites with Tunable
  Optoelectronic Properties Due to Reversible Surface Relaxation}},\ }\href
  {https://doi.org/10.1038/s41563-018-0164-8} {\bibfield  {journal} {\bibinfo
  {journal} {Nat. Mater.}\ }\textbf {\bibinfo {volume} {17}},\ \bibinfo {pages}
  {908} (\bibinfo {year} {2018})}\BibitemShut {NoStop}%
\bibitem [{\citenamefont {Shi}\ \emph {et~al.}(2018)\citenamefont {Shi},
  \citenamefont {Gao}, \citenamefont {Finkenauer}, \citenamefont {Akriti},
  \citenamefont {Coffey},\ and\ \citenamefont {Dou}}]{SGplanar18}%
  \BibitemOpen
  \bibfield  {author} {\bibinfo {author} {\bibfnamefont {E.}~\bibnamefont
  {Shi}}, \bibinfo {author} {\bibfnamefont {Y.}~\bibnamefont {Gao}}, \bibinfo
  {author} {\bibfnamefont {B.~P.}\ \bibnamefont {Finkenauer}}, \bibinfo
  {author} {\bibnamefont {Akriti}}, \bibinfo {author} {\bibfnamefont {A.~H.}\
  \bibnamefont {Coffey}},\ and\ \bibinfo {author} {\bibfnamefont
  {L.}~\bibnamefont {Dou}},\ }\bibfield  {title} {\bibinfo {title}
  {{Two-Dimensional Halide Perovskite Nanomaterials and Heterostructures}},\
  }\href {https://doi.org/10.1039/C7CS00886D} {\bibfield  {journal} {\bibinfo
  {journal} {Chem. Soc. Rev.}\ }\textbf {\bibinfo {volume} {47}},\ \bibinfo
  {pages} {6046} (\bibinfo {year} {2018})}\BibitemShut {NoStop}%
\bibitem [{\citenamefont {Cao}\ \emph {et~al.}(2017)\citenamefont {Cao},
  \citenamefont {Stoumpos}, \citenamefont {Yokoyama}, \citenamefont {Logsdon},
  \citenamefont {Song}, \citenamefont {Farha}, \citenamefont {Wasielewski},
  \citenamefont {Hupp},\ and\ \citenamefont {Kanatzidis}}]{CSpbfree17}%
  \BibitemOpen
  \bibfield  {author} {\bibinfo {author} {\bibfnamefont {D.~H.}\ \bibnamefont
  {Cao}}, \bibinfo {author} {\bibfnamefont {C.~C.}\ \bibnamefont {Stoumpos}},
  \bibinfo {author} {\bibfnamefont {T.}~\bibnamefont {Yokoyama}}, \bibinfo
  {author} {\bibfnamefont {J.~L.}\ \bibnamefont {Logsdon}}, \bibinfo {author}
  {\bibfnamefont {T.-B.}\ \bibnamefont {Song}}, \bibinfo {author}
  {\bibfnamefont {O.~K.}\ \bibnamefont {Farha}}, \bibinfo {author}
  {\bibfnamefont {M.~R.}\ \bibnamefont {Wasielewski}}, \bibinfo {author}
  {\bibfnamefont {J.~T.}\ \bibnamefont {Hupp}},\ and\ \bibinfo {author}
  {\bibfnamefont {M.~G.}\ \bibnamefont {Kanatzidis}},\ }\bibfield  {title}
  {\bibinfo {title} {{Thin Films and Solar Cells Based on Semiconducting
  Two-Dimensional Ruddlesden-Popper
  (CH$_{3}$(CH$_{2}$)$_{3}$NH$_{3}$)$_{2}$(CH$_{3}$NH$_{3}$)$_{n-1}$Sn$_{n}$I$_{3n+1}$
  Perovskites}},\ }\href {https://doi.org/10.1021/acsenergylett.7b00202}
  {\bibfield  {journal} {\bibinfo  {journal} {ACS Energy Lett.}\ }\textbf
  {\bibinfo {volume} {2}},\ \bibinfo {pages} {982} (\bibinfo {year}
  {2017})}\BibitemShut {NoStop}%
\bibitem [{\citenamefont {Cheng}\ \emph {et~al.}(2017)\citenamefont {Cheng},
  \citenamefont {Wu}, \citenamefont {Zhang}, \citenamefont {Li}, \citenamefont
  {Liu}, \citenamefont {Jiang}, \citenamefont {Mao}, \citenamefont {Lu},
  \citenamefont {Deng},\ and\ \citenamefont {Han}}]{CWpbfree17}%
  \BibitemOpen
  \bibfield  {author} {\bibinfo {author} {\bibfnamefont {P.}~\bibnamefont
  {Cheng}}, \bibinfo {author} {\bibfnamefont {T.}~\bibnamefont {Wu}}, \bibinfo
  {author} {\bibfnamefont {J.}~\bibnamefont {Zhang}}, \bibinfo {author}
  {\bibfnamefont {Y.}~\bibnamefont {Li}}, \bibinfo {author} {\bibfnamefont
  {J.}~\bibnamefont {Liu}}, \bibinfo {author} {\bibfnamefont {L.}~\bibnamefont
  {Jiang}}, \bibinfo {author} {\bibfnamefont {X.}~\bibnamefont {Mao}}, \bibinfo
  {author} {\bibfnamefont {R.-F.}\ \bibnamefont {Lu}}, \bibinfo {author}
  {\bibfnamefont {W.-Q.}\ \bibnamefont {Deng}},\ and\ \bibinfo {author}
  {\bibfnamefont {K.}~\bibnamefont {Han}},\ }\bibfield  {title} {\bibinfo
  {title} {{(C$_{6}$H$_{5}$C$_{2}$H$_{4}$NH$_{3}$)$_{2}$GeI$_{4}$: A Layered
  Two-Dimensional Perovskite with Potential for Photovoltaic Applications}},\
  }\href {https://doi.org/10.1021/acs.jpclett.7b01985} {\bibfield  {journal}
  {\bibinfo  {journal} {J. Phys. Chem. Lett.}\ }\textbf {\bibinfo {volume}
  {8}},\ \bibinfo {pages} {4402} (\bibinfo {year} {2017})}\BibitemShut
  {NoStop}%
\bibitem [{\citenamefont {McClure}\ \emph {et~al.}(2020)\citenamefont
  {McClure}, \citenamefont {McCormick},\ and\ \citenamefont
  {Woodward}}]{MMpbfree20}%
  \BibitemOpen
  \bibfield  {author} {\bibinfo {author} {\bibfnamefont {E.~T.}\ \bibnamefont
  {McClure}}, \bibinfo {author} {\bibfnamefont {A.~P.}\ \bibnamefont
  {McCormick}},\ and\ \bibinfo {author} {\bibfnamefont {P.~M.}\ \bibnamefont
  {Woodward}},\ }\bibfield  {title} {\bibinfo {title} {{Four Lead-Free Layered
  Double Perovskites with the $n$ = 1 Ruddlesden-Popper Structure}},\ }\href
  {https://doi.org/10.1021/acs.inorgchem.0c00009} {\bibfield  {journal}
  {\bibinfo  {journal} {Inorg. Chem.}\ }\textbf {\bibinfo {volume} {59}},\
  \bibinfo {pages} {6010} (\bibinfo {year} {2020})}\BibitemShut {NoStop}%
\bibitem [{\citenamefont {Guo}\ \emph {et~al.}(2016)\citenamefont {Guo},
  \citenamefont {Wu}, \citenamefont {Zhu}, \citenamefont {Zhu},\ and\
  \citenamefont {Huang}}]{GWexf16}%
  \BibitemOpen
  \bibfield  {author} {\bibinfo {author} {\bibfnamefont {Z.}~\bibnamefont
  {Guo}}, \bibinfo {author} {\bibfnamefont {X.}~\bibnamefont {Wu}}, \bibinfo
  {author} {\bibfnamefont {T.}~\bibnamefont {Zhu}}, \bibinfo {author}
  {\bibfnamefont {X.}~\bibnamefont {Zhu}},\ and\ \bibinfo {author}
  {\bibfnamefont {L.}~\bibnamefont {Huang}},\ }\bibfield  {title} {\bibinfo
  {title} {{Electron-Phonon Scattering in Atomically Thin 2D Perovskites}},\
  }\href {https://doi.org/10.1021/acsnano.6b04265} {\bibfield  {journal}
  {\bibinfo  {journal} {ACS Nano}\ }\textbf {\bibinfo {volume} {10}},\ \bibinfo
  {pages} {9992} (\bibinfo {year} {2016})}\BibitemShut {NoStop}%
\bibitem [{\citenamefont {Yaffe}\ \emph {et~al.}(2015)\citenamefont {Yaffe},
  \citenamefont {Chernikov}, \citenamefont {Norman}, \citenamefont {Zhong},
  \citenamefont {Velauthapillai}, \citenamefont {van~der Zande}, \citenamefont
  {Owen},\ and\ \citenamefont {Heinz}}]{YCexf15}%
  \BibitemOpen
  \bibfield  {author} {\bibinfo {author} {\bibfnamefont {O.}~\bibnamefont
  {Yaffe}}, \bibinfo {author} {\bibfnamefont {A.}~\bibnamefont {Chernikov}},
  \bibinfo {author} {\bibfnamefont {Z.~M.}\ \bibnamefont {Norman}}, \bibinfo
  {author} {\bibfnamefont {Y.}~\bibnamefont {Zhong}}, \bibinfo {author}
  {\bibfnamefont {A.}~\bibnamefont {Velauthapillai}}, \bibinfo {author}
  {\bibfnamefont {A.}~\bibnamefont {van~der Zande}}, \bibinfo {author}
  {\bibfnamefont {J.~S.}\ \bibnamefont {Owen}},\ and\ \bibinfo {author}
  {\bibfnamefont {T.~F.}\ \bibnamefont {Heinz}},\ }\bibfield  {title} {\bibinfo
  {title} {{Excitons in Ultrathin Organic-Inorganic Perovskite Crystals}},\
  }\href {https://doi.org/10.1103/PhysRevB.92.045414} {\bibfield  {journal}
  {\bibinfo  {journal} {Phys. Rev. B}\ }\textbf {\bibinfo {volume} {92}},\
  \bibinfo {pages} {045414} (\bibinfo {year} {2015})}\BibitemShut {NoStop}%
\bibitem [{\citenamefont {Niu}\ \emph {et~al.}(2014)\citenamefont {Niu},
  \citenamefont {Eiden}, \citenamefont {Vijaya~Prakash},\ and\ \citenamefont
  {Baumberg}}]{NEexf14}%
  \BibitemOpen
  \bibfield  {author} {\bibinfo {author} {\bibfnamefont {W.}~\bibnamefont
  {Niu}}, \bibinfo {author} {\bibfnamefont {A.}~\bibnamefont {Eiden}}, \bibinfo
  {author} {\bibfnamefont {G.}~\bibnamefont {Vijaya~Prakash}},\ and\ \bibinfo
  {author} {\bibfnamefont {J.~J.}\ \bibnamefont {Baumberg}},\ }\bibfield
  {title} {\bibinfo {title} {{Exfoliation of Self-Assembled 2D
  Organic-Inorganic Perovskite Semiconductors}},\ }\href
  {https://doi.org/10.1063/1.4874846} {\bibfield  {journal} {\bibinfo
  {journal} {Appl. Phys. Lett.}\ }\textbf {\bibinfo {volume} {104}},\ \bibinfo
  {pages} {171111} (\bibinfo {year} {2014})}\BibitemShut {NoStop}%
\bibitem [{\citenamefont {Gao}\ \emph {et~al.}(2019)\citenamefont {Gao},
  \citenamefont {Shi}, \citenamefont {Deng}, \citenamefont {Shiring},
  \citenamefont {Snaider}, \citenamefont {Liang}, \citenamefont {Yuan},
  \citenamefont {Song}, \citenamefont {Janke}, \citenamefont
  {Liebman-Pel{\'a}ez}, \citenamefont {Yoo}, \citenamefont {Zeller},
  \citenamefont {Boudouris}, \citenamefont {Liao}, \citenamefont {Zhu},
  \citenamefont {Blum}, \citenamefont {Yu}, \citenamefont {Savoie},
  \citenamefont {Huang},\ and\ \citenamefont {Dou}}]{GSsolv19}%
  \BibitemOpen
  \bibfield  {author} {\bibinfo {author} {\bibfnamefont {Y.}~\bibnamefont
  {Gao}}, \bibinfo {author} {\bibfnamefont {E.}~\bibnamefont {Shi}}, \bibinfo
  {author} {\bibfnamefont {S.}~\bibnamefont {Deng}}, \bibinfo {author}
  {\bibfnamefont {S.~B.}\ \bibnamefont {Shiring}}, \bibinfo {author}
  {\bibfnamefont {J.~M.}\ \bibnamefont {Snaider}}, \bibinfo {author}
  {\bibfnamefont {C.}~\bibnamefont {Liang}}, \bibinfo {author} {\bibfnamefont
  {B.}~\bibnamefont {Yuan}}, \bibinfo {author} {\bibfnamefont {R.}~\bibnamefont
  {Song}}, \bibinfo {author} {\bibfnamefont {S.~M.}\ \bibnamefont {Janke}},
  \bibinfo {author} {\bibfnamefont {A.}~\bibnamefont {Liebman-Pel{\'a}ez}},
  \bibinfo {author} {\bibfnamefont {P.}~\bibnamefont {Yoo}}, \bibinfo {author}
  {\bibfnamefont {M.}~\bibnamefont {Zeller}}, \bibinfo {author} {\bibfnamefont
  {B.~W.}\ \bibnamefont {Boudouris}}, \bibinfo {author} {\bibfnamefont
  {P.}~\bibnamefont {Liao}}, \bibinfo {author} {\bibfnamefont {C.}~\bibnamefont
  {Zhu}}, \bibinfo {author} {\bibfnamefont {V.}~\bibnamefont {Blum}}, \bibinfo
  {author} {\bibfnamefont {Y.}~\bibnamefont {Yu}}, \bibinfo {author}
  {\bibfnamefont {B.~M.}\ \bibnamefont {Savoie}}, \bibinfo {author}
  {\bibfnamefont {L.}~\bibnamefont {Huang}},\ and\ \bibinfo {author}
  {\bibfnamefont {L.}~\bibnamefont {Dou}},\ }\bibfield  {title} {\bibinfo
  {title} {{Molecular Engineering of Organic-Inorganic Hybrid Perovskites
  Quantum Wells}},\ }\href {https://doi.org/10.1038/s41557-019-0354-2}
  {\bibfield  {journal} {\bibinfo  {journal} {Nat. Chem.}\ }\textbf {\bibinfo
  {volume} {11}},\ \bibinfo {pages} {1151} (\bibinfo {year}
  {2019})}\BibitemShut {NoStop}%
\bibitem [{\citenamefont {Dou}\ \emph {et~al.}(2015)\citenamefont {Dou},
  \citenamefont {Wong}, \citenamefont {Yu}, \citenamefont {Lai}, \citenamefont
  {Kornienko}, \citenamefont {Eaton}, \citenamefont {Fu}, \citenamefont
  {Bischak}, \citenamefont {Ma}, \citenamefont {Ding}, \citenamefont
  {Ginsberg}, \citenamefont {Wang}, \citenamefont {Alivisatos},\ and\
  \citenamefont {Yang}}]{DWsolv15}%
  \BibitemOpen
  \bibfield  {author} {\bibinfo {author} {\bibfnamefont {L.}~\bibnamefont
  {Dou}}, \bibinfo {author} {\bibfnamefont {A.~B.}\ \bibnamefont {Wong}},
  \bibinfo {author} {\bibfnamefont {Y.}~\bibnamefont {Yu}}, \bibinfo {author}
  {\bibfnamefont {M.}~\bibnamefont {Lai}}, \bibinfo {author} {\bibfnamefont
  {N.}~\bibnamefont {Kornienko}}, \bibinfo {author} {\bibfnamefont {S.~W.}\
  \bibnamefont {Eaton}}, \bibinfo {author} {\bibfnamefont {A.}~\bibnamefont
  {Fu}}, \bibinfo {author} {\bibfnamefont {C.~G.}\ \bibnamefont {Bischak}},
  \bibinfo {author} {\bibfnamefont {J.}~\bibnamefont {Ma}}, \bibinfo {author}
  {\bibfnamefont {T.}~\bibnamefont {Ding}}, \bibinfo {author} {\bibfnamefont
  {N.~S.}\ \bibnamefont {Ginsberg}}, \bibinfo {author} {\bibfnamefont {L.-W.}\
  \bibnamefont {Wang}}, \bibinfo {author} {\bibfnamefont {A.~P.}\ \bibnamefont
  {Alivisatos}},\ and\ \bibinfo {author} {\bibfnamefont {P.}~\bibnamefont
  {Yang}},\ }\bibfield  {title} {\bibinfo {title} {{Atomically Thin
  Two-Dimensional Organic-Inorganic Hybrid Perovskites}},\ }\href
  {https://doi.org/10.1126/science.aac7660} {\bibfield  {journal} {\bibinfo
  {journal} {Science}\ }\textbf {\bibinfo {volume} {349}},\ \bibinfo {pages}
  {1518} (\bibinfo {year} {2015})}\BibitemShut {NoStop}%
\bibitem [{\citenamefont {Tu}\ \emph {et~al.}(2018)\citenamefont {Tu},
  \citenamefont {Spanopoulos}, \citenamefont {Yasaei}, \citenamefont
  {Stoumpos}, \citenamefont {Kanatzidis}, \citenamefont {Shekhawat},\ and\
  \citenamefont {Dravid}}]{TSultra2d18}%
  \BibitemOpen
  \bibfield  {author} {\bibinfo {author} {\bibfnamefont {Q.}~\bibnamefont
  {Tu}}, \bibinfo {author} {\bibfnamefont {I.}~\bibnamefont {Spanopoulos}},
  \bibinfo {author} {\bibfnamefont {P.}~\bibnamefont {Yasaei}}, \bibinfo
  {author} {\bibfnamefont {C.~C.}\ \bibnamefont {Stoumpos}}, \bibinfo {author}
  {\bibfnamefont {M.~G.}\ \bibnamefont {Kanatzidis}}, \bibinfo {author}
  {\bibfnamefont {G.~S.}\ \bibnamefont {Shekhawat}},\ and\ \bibinfo {author}
  {\bibfnamefont {V.~P.}\ \bibnamefont {Dravid}},\ }\bibfield  {title}
  {\bibinfo {title} {{Stretching and Breaking of Ultrathin 2D Hybrid
  Organic-Inorganic Perovskites}},\ }\href
  {https://doi.org/10.1021/acsnano.8b05623} {\bibfield  {journal} {\bibinfo
  {journal} {ACS Nano}\ }\textbf {\bibinfo {volume} {12}},\ \bibinfo {pages}
  {10347} (\bibinfo {year} {2018})}\BibitemShut {NoStop}%
\bibitem [{\citenamefont {Qi}\ \emph {et~al.}(2018)\citenamefont {Qi},
  \citenamefont {Zhang}, \citenamefont {Ou}, \citenamefont {Ha}, \citenamefont
  {Qiu}, \citenamefont {Zhang}, \citenamefont {Cheng}, \citenamefont {Xiong},\
  and\ \citenamefont {Bao}}]{QZultra2d18}%
  \BibitemOpen
  \bibfield  {author} {\bibinfo {author} {\bibfnamefont {X.}~\bibnamefont
  {Qi}}, \bibinfo {author} {\bibfnamefont {Y.}~\bibnamefont {Zhang}}, \bibinfo
  {author} {\bibfnamefont {Q.}~\bibnamefont {Ou}}, \bibinfo {author}
  {\bibfnamefont {S.~T.}\ \bibnamefont {Ha}}, \bibinfo {author} {\bibfnamefont
  {C.-W.}\ \bibnamefont {Qiu}}, \bibinfo {author} {\bibfnamefont
  {H.}~\bibnamefont {Zhang}}, \bibinfo {author} {\bibfnamefont {Y.-B.}\
  \bibnamefont {Cheng}}, \bibinfo {author} {\bibfnamefont {Q.}~\bibnamefont
  {Xiong}},\ and\ \bibinfo {author} {\bibfnamefont {Q.}~\bibnamefont {Bao}},\
  }\bibfield  {title} {\bibinfo {title} {{Photonics and Optoelectronics of 2D
  Metal-Halide Perovskites}},\ }\href
  {https://doi.org/https://doi.org/10.1002/smll.201800682} {\bibfield
  {journal} {\bibinfo  {journal} {Small}\ }\textbf {\bibinfo {volume} {14}},\
  \bibinfo {pages} {1800682} (\bibinfo {year} {2018})}\BibitemShut {NoStop}%
\bibitem [{\citenamefont {Ou}\ \emph {et~al.}(2018)\citenamefont {Ou},
  \citenamefont {Zhang}, \citenamefont {Wang}, \citenamefont {Yuwono},
  \citenamefont {Wang}, \citenamefont {Dai}, \citenamefont {Li}, \citenamefont
  {Zheng}, \citenamefont {Xu}, \citenamefont {Qi}, \citenamefont {Duhm},
  \citenamefont {Medhekar}, \citenamefont {Zhang},\ and\ \citenamefont
  {Bao}}]{OZlowdp18}%
  \BibitemOpen
  \bibfield  {author} {\bibinfo {author} {\bibfnamefont {Q.}~\bibnamefont
  {Ou}}, \bibinfo {author} {\bibfnamefont {Y.}~\bibnamefont {Zhang}}, \bibinfo
  {author} {\bibfnamefont {Z.}~\bibnamefont {Wang}}, \bibinfo {author}
  {\bibfnamefont {J.~A.}\ \bibnamefont {Yuwono}}, \bibinfo {author}
  {\bibfnamefont {R.}~\bibnamefont {Wang}}, \bibinfo {author} {\bibfnamefont
  {Z.}~\bibnamefont {Dai}}, \bibinfo {author} {\bibfnamefont {W.}~\bibnamefont
  {Li}}, \bibinfo {author} {\bibfnamefont {C.}~\bibnamefont {Zheng}}, \bibinfo
  {author} {\bibfnamefont {Z.-Q.}\ \bibnamefont {Xu}}, \bibinfo {author}
  {\bibfnamefont {X.}~\bibnamefont {Qi}}, \bibinfo {author} {\bibfnamefont
  {S.}~\bibnamefont {Duhm}}, \bibinfo {author} {\bibfnamefont {N.~V.}\
  \bibnamefont {Medhekar}}, \bibinfo {author} {\bibfnamefont {H.}~\bibnamefont
  {Zhang}},\ and\ \bibinfo {author} {\bibfnamefont {Q.}~\bibnamefont {Bao}},\
  }\bibfield  {title} {\bibinfo {title} {{Strong Depletion in Hybrid Perovskite
  \textit{p-n} Junctions Induced by Local Electronic Doping}},\ }\href
  {https://doi.org/https://doi.org/10.1002/adma.201705792} {\bibfield
  {journal} {\bibinfo  {journal} {Adv. Mater.}\ }\textbf {\bibinfo {volume}
  {30}},\ \bibinfo {pages} {1705792} (\bibinfo {year} {2018})}\BibitemShut
  {NoStop}%
\bibitem [{\citenamefont {Chen}\ \emph {et~al.}(2019)\citenamefont {Chen},
  \citenamefont {Zhong}, \citenamefont {Chen}, \citenamefont {Sang},
  \citenamefont {Wang}, \citenamefont {Yang}, \citenamefont {Liu},
  \citenamefont {Zhang},\ and\ \citenamefont {Zhang}}]{CZlowdp19}%
  \BibitemOpen
  \bibfield  {author} {\bibinfo {author} {\bibfnamefont {K.}~\bibnamefont
  {Chen}}, \bibinfo {author} {\bibfnamefont {Q.}~\bibnamefont {Zhong}},
  \bibinfo {author} {\bibfnamefont {W.}~\bibnamefont {Chen}}, \bibinfo {author}
  {\bibfnamefont {B.}~\bibnamefont {Sang}}, \bibinfo {author} {\bibfnamefont
  {Y.}~\bibnamefont {Wang}}, \bibinfo {author} {\bibfnamefont {T.}~\bibnamefont
  {Yang}}, \bibinfo {author} {\bibfnamefont {Y.}~\bibnamefont {Liu}}, \bibinfo
  {author} {\bibfnamefont {Y.}~\bibnamefont {Zhang}},\ and\ \bibinfo {author}
  {\bibfnamefont {H.}~\bibnamefont {Zhang}},\ }\bibfield  {title} {\bibinfo
  {title} {{Short-Chain Ligand-Passivated Stable $\alpha$-CsPbI$_3$ Quantum Dot
  for All-Inorganic Perovskite Solar Cells}},\ }\href
  {https://doi.org/https://doi.org/10.1002/adfm.201900991} {\bibfield
  {journal} {\bibinfo  {journal} {Adv. Funct. Mater.}\ }\textbf {\bibinfo
  {volume} {29}},\ \bibinfo {pages} {1900991} (\bibinfo {year}
  {2019})}\BibitemShut {NoStop}%
\bibitem [{\citenamefont {Blancon}\ \emph {et~al.}(2017)\citenamefont
  {Blancon}, \citenamefont {Tsai}, \citenamefont {Nie}, \citenamefont
  {Stoumpos}, \citenamefont {Pedesseau}, \citenamefont {Katan}, \citenamefont
  {Kepenekian}, \citenamefont {Soe}, \citenamefont {Appavoo}, \citenamefont
  {Sfeir}, \citenamefont {Tretiak}, \citenamefont {Ajayan}, \citenamefont
  {Kanatzidis}, \citenamefont {Even}, \citenamefont {Crochet},\ and\
  \citenamefont {Mohite}}]{BTles17}%
  \BibitemOpen
  \bibfield  {author} {\bibinfo {author} {\bibfnamefont {J.-C.}\ \bibnamefont
  {Blancon}}, \bibinfo {author} {\bibfnamefont {H.}~\bibnamefont {Tsai}},
  \bibinfo {author} {\bibfnamefont {W.}~\bibnamefont {Nie}}, \bibinfo {author}
  {\bibfnamefont {C.~C.}\ \bibnamefont {Stoumpos}}, \bibinfo {author}
  {\bibfnamefont {L.}~\bibnamefont {Pedesseau}}, \bibinfo {author}
  {\bibfnamefont {C.}~\bibnamefont {Katan}}, \bibinfo {author} {\bibfnamefont
  {M.}~\bibnamefont {Kepenekian}}, \bibinfo {author} {\bibfnamefont {C.~M.~M.}\
  \bibnamefont {Soe}}, \bibinfo {author} {\bibfnamefont {K.}~\bibnamefont
  {Appavoo}}, \bibinfo {author} {\bibfnamefont {M.~Y.}\ \bibnamefont {Sfeir}},
  \bibinfo {author} {\bibfnamefont {S.}~\bibnamefont {Tretiak}}, \bibinfo
  {author} {\bibfnamefont {P.~M.}\ \bibnamefont {Ajayan}}, \bibinfo {author}
  {\bibfnamefont {M.~G.}\ \bibnamefont {Kanatzidis}}, \bibinfo {author}
  {\bibfnamefont {J.}~\bibnamefont {Even}}, \bibinfo {author} {\bibfnamefont
  {J.~J.}\ \bibnamefont {Crochet}},\ and\ \bibinfo {author} {\bibfnamefont
  {A.~D.}\ \bibnamefont {Mohite}},\ }\bibfield  {title} {\bibinfo {title}
  {{Extremely Efficient Internal Exciton Dissociation through Edge States in
  Layered 2D Perovskites}},\ }\href {https://doi.org/10.1126/science.aal4211}
  {\bibfield  {journal} {\bibinfo  {journal} {Science}\ }\textbf {\bibinfo
  {volume} {355}},\ \bibinfo {pages} {1288} (\bibinfo {year}
  {2017})}\BibitemShut {NoStop}%
\bibitem [{\citenamefont {Qin}\ \emph {et~al.}(2020)\citenamefont {Qin},
  \citenamefont {Dai}, \citenamefont {Gajjela}, \citenamefont {Wang},
  \citenamefont {Hadjiev}, \citenamefont {Yang}, \citenamefont {Li},
  \citenamefont {Zhong}, \citenamefont {Tang}, \citenamefont {Yao},
  \citenamefont {Guloy}, \citenamefont {Reddy}, \citenamefont {Mayerich},
  \citenamefont {Deng}, \citenamefont {Yu}, \citenamefont {Feng}, \citenamefont
  {Calderon}, \citenamefont {Robles~Hernandez}, \citenamefont {Wang},\ and\
  \citenamefont {Bao}}]{QDhetedge20}%
  \BibitemOpen
  \bibfield  {author} {\bibinfo {author} {\bibfnamefont {Z.}~\bibnamefont
  {Qin}}, \bibinfo {author} {\bibfnamefont {S.}~\bibnamefont {Dai}}, \bibinfo
  {author} {\bibfnamefont {C.~C.}\ \bibnamefont {Gajjela}}, \bibinfo {author}
  {\bibfnamefont {C.}~\bibnamefont {Wang}}, \bibinfo {author} {\bibfnamefont
  {V.~G.}\ \bibnamefont {Hadjiev}}, \bibinfo {author} {\bibfnamefont
  {G.}~\bibnamefont {Yang}}, \bibinfo {author} {\bibfnamefont {J.}~\bibnamefont
  {Li}}, \bibinfo {author} {\bibfnamefont {X.}~\bibnamefont {Zhong}}, \bibinfo
  {author} {\bibfnamefont {Z.}~\bibnamefont {Tang}}, \bibinfo {author}
  {\bibfnamefont {Y.}~\bibnamefont {Yao}}, \bibinfo {author} {\bibfnamefont
  {A.~M.}\ \bibnamefont {Guloy}}, \bibinfo {author} {\bibfnamefont
  {R.}~\bibnamefont {Reddy}}, \bibinfo {author} {\bibfnamefont
  {D.}~\bibnamefont {Mayerich}}, \bibinfo {author} {\bibfnamefont
  {L.}~\bibnamefont {Deng}}, \bibinfo {author} {\bibfnamefont {Q.}~\bibnamefont
  {Yu}}, \bibinfo {author} {\bibfnamefont {G.}~\bibnamefont {Feng}}, \bibinfo
  {author} {\bibfnamefont {H.~A.}\ \bibnamefont {Calderon}}, \bibinfo {author}
  {\bibfnamefont {F.~C.}\ \bibnamefont {Robles~Hernandez}}, \bibinfo {author}
  {\bibfnamefont {Z.~M.}\ \bibnamefont {Wang}},\ and\ \bibinfo {author}
  {\bibfnamefont {J.}~\bibnamefont {Bao}},\ }\bibfield  {title} {\bibinfo
  {title} {{Spontaneous Formation of 2D/3D Heterostructures on the Edges of 2D
  Ruddlesden-Popper Hybrid Perovskite Crystals}},\ }\href
  {https://doi.org/10.1021/acs.chemmater.0c00419} {\bibfield  {journal}
  {\bibinfo  {journal} {Chem. Mater.}\ }\textbf {\bibinfo {volume} {32}},\
  \bibinfo {pages} {5009} (\bibinfo {year} {2020})}\BibitemShut {NoStop}%
\bibitem [{\citenamefont {Zhao}\ \emph {et~al.}(2019)\citenamefont {Zhao},
  \citenamefont {Tian}, \citenamefont {Leng}, \citenamefont {Zhao},\ and\
  \citenamefont {Jin}}]{ZTrevcon19}%
  \BibitemOpen
  \bibfield  {author} {\bibinfo {author} {\bibfnamefont {C.}~\bibnamefont
  {Zhao}}, \bibinfo {author} {\bibfnamefont {W.}~\bibnamefont {Tian}}, \bibinfo
  {author} {\bibfnamefont {J.}~\bibnamefont {Leng}}, \bibinfo {author}
  {\bibfnamefont {Y.}~\bibnamefont {Zhao}},\ and\ \bibinfo {author}
  {\bibfnamefont {S.}~\bibnamefont {Jin}},\ }\bibfield  {title} {\bibinfo
  {title} {{Controlling the Property of Edges in Layered 2D Perovskite Single
  Crystals}},\ }\href {https://doi.org/10.1021/acs.jpclett.9b01193} {\bibfield
  {journal} {\bibinfo  {journal} {J. Phys. Chem. Lett.}\ }\textbf {\bibinfo
  {volume} {10}},\ \bibinfo {pages} {3950} (\bibinfo {year}
  {2019})}\BibitemShut {NoStop}%
\bibitem [{\citenamefont {Shi}\ \emph {et~al.}(2019)\citenamefont {Shi},
  \citenamefont {Deng}, \citenamefont {Yuan}, \citenamefont {Gao},
  \citenamefont {Akriti}, \citenamefont {Yuan}, \citenamefont {Davis},
  \citenamefont {Zemlyanov}, \citenamefont {Yu}, \citenamefont {Huang},\ and\
  \citenamefont {Dou}}]{SDhumid19}%
  \BibitemOpen
  \bibfield  {author} {\bibinfo {author} {\bibfnamefont {E.}~\bibnamefont
  {Shi}}, \bibinfo {author} {\bibfnamefont {S.}~\bibnamefont {Deng}}, \bibinfo
  {author} {\bibfnamefont {B.}~\bibnamefont {Yuan}}, \bibinfo {author}
  {\bibfnamefont {Y.}~\bibnamefont {Gao}}, \bibinfo {author} {\bibnamefont
  {Akriti}}, \bibinfo {author} {\bibfnamefont {L.}~\bibnamefont {Yuan}},
  \bibinfo {author} {\bibfnamefont {C.~S.}\ \bibnamefont {Davis}}, \bibinfo
  {author} {\bibfnamefont {D.}~\bibnamefont {Zemlyanov}}, \bibinfo {author}
  {\bibfnamefont {Y.}~\bibnamefont {Yu}}, \bibinfo {author} {\bibfnamefont
  {L.}~\bibnamefont {Huang}},\ and\ \bibinfo {author} {\bibfnamefont
  {L.}~\bibnamefont {Dou}},\ }\bibfield  {title} {\bibinfo {title} {{Extrinsic
  and Dynamic Edge States of Two-Dimensional Lead Halide Perovskites}},\ }\href
  {https://doi.org/10.1021/acsnano.8b07631} {\bibfield  {journal} {\bibinfo
  {journal} {ACS Nano}\ }\textbf {\bibinfo {volume} {13}},\ \bibinfo {pages}
  {1635} (\bibinfo {year} {2019})}\BibitemShut {NoStop}%
\bibitem [{\citenamefont {Guo}\ \emph {et~al.}(2017)\citenamefont {Guo},
  \citenamefont {Yang}, \citenamefont {Leng}, \citenamefont {Wang},
  \citenamefont {Dong}, \citenamefont {Liu},\ and\ \citenamefont
  {Li}}]{GYedgegrtmd17}%
  \BibitemOpen
  \bibfield  {author} {\bibinfo {author} {\bibfnamefont {M.}~\bibnamefont
  {Guo}}, \bibinfo {author} {\bibfnamefont {Y.}~\bibnamefont {Yang}}, \bibinfo
  {author} {\bibfnamefont {Y.}~\bibnamefont {Leng}}, \bibinfo {author}
  {\bibfnamefont {L.}~\bibnamefont {Wang}}, \bibinfo {author} {\bibfnamefont
  {H.}~\bibnamefont {Dong}}, \bibinfo {author} {\bibfnamefont {H.}~\bibnamefont
  {Liu}},\ and\ \bibinfo {author} {\bibfnamefont {W.}~\bibnamefont {Li}},\
  }\bibfield  {title} {\bibinfo {title} {{Edge Dominated Electronic Properties
  of MoS$_2$/Graphene Hybrid 2D Materials: Edge State, Electron Coupling and
  Work Function}},\ }\href {https://doi.org/10.1039/C7TC00816C} {\bibfield
  {journal} {\bibinfo  {journal} {J. Mater. Chem. C}\ }\textbf {\bibinfo
  {volume} {5}},\ \bibinfo {pages} {4845} (\bibinfo {year} {2017})}\BibitemShut
  {NoStop}%
\bibitem [{\citenamefont {Huang}\ \emph {et~al.}(2009)\citenamefont {Huang},
  \citenamefont {Liu}, \citenamefont {Su}, \citenamefont {Wu}, \citenamefont
  {Duan}, \citenamefont {Gu},\ and\ \citenamefont {Liu}}]{HLedgegr09}%
  \BibitemOpen
  \bibfield  {author} {\bibinfo {author} {\bibfnamefont {B.}~\bibnamefont
  {Huang}}, \bibinfo {author} {\bibfnamefont {M.}~\bibnamefont {Liu}}, \bibinfo
  {author} {\bibfnamefont {N.}~\bibnamefont {Su}}, \bibinfo {author}
  {\bibfnamefont {J.}~\bibnamefont {Wu}}, \bibinfo {author} {\bibfnamefont
  {W.}~\bibnamefont {Duan}}, \bibinfo {author} {\bibfnamefont {B.-l.}\
  \bibnamefont {Gu}},\ and\ \bibinfo {author} {\bibfnamefont {F.}~\bibnamefont
  {Liu}},\ }\bibfield  {title} {\bibinfo {title} {{Quantum Manifestations of
  Graphene Edge Stress and Edge Instability: A First-Principles Study}},\
  }\href {https://doi.org/10.1103/PhysRevLett.102.166404} {\bibfield  {journal}
  {\bibinfo  {journal} {Phys. Rev. Lett.}\ }\textbf {\bibinfo {volume} {102}},\
  \bibinfo {pages} {166404} (\bibinfo {year} {2009})}\BibitemShut {NoStop}%
\bibitem [{\citenamefont {Shenoy}\ \emph {et~al.}(2008)\citenamefont {Shenoy},
  \citenamefont {Reddy}, \citenamefont {Ramasubramaniam},\ and\ \citenamefont
  {Zhang}}]{SRedgegr08}%
  \BibitemOpen
  \bibfield  {author} {\bibinfo {author} {\bibfnamefont {V.~B.}\ \bibnamefont
  {Shenoy}}, \bibinfo {author} {\bibfnamefont {C.~D.}\ \bibnamefont {Reddy}},
  \bibinfo {author} {\bibfnamefont {A.}~\bibnamefont {Ramasubramaniam}},\ and\
  \bibinfo {author} {\bibfnamefont {Y.~W.}\ \bibnamefont {Zhang}},\ }\bibfield
  {title} {\bibinfo {title} {{Edge-Stress-Induced Warping of Graphene Sheets
  and Nanoribbons}},\ }\href {https://doi.org/10.1103/PhysRevLett.101.245501}
  {\bibfield  {journal} {\bibinfo  {journal} {Phys. Rev. Lett.}\ }\textbf
  {\bibinfo {volume} {101}},\ \bibinfo {pages} {245501} (\bibinfo {year}
  {2008})}\BibitemShut {NoStop}%
\bibitem [{\citenamefont {Singh}\ \emph {et~al.}(2020)\citenamefont {Singh},
  \citenamefont {Panda}, \citenamefont {Khossossi}, \citenamefont {Mishra},
  \citenamefont {Ainane},\ and\ \citenamefont {Ahuja}}]{SPedgetmd20}%
  \BibitemOpen
  \bibfield  {author} {\bibinfo {author} {\bibfnamefont {D.}~\bibnamefont
  {Singh}}, \bibinfo {author} {\bibfnamefont {P.~K.}\ \bibnamefont {Panda}},
  \bibinfo {author} {\bibfnamefont {N.}~\bibnamefont {Khossossi}}, \bibinfo
  {author} {\bibfnamefont {Y.~K.}\ \bibnamefont {Mishra}}, \bibinfo {author}
  {\bibfnamefont {A.}~\bibnamefont {Ainane}},\ and\ \bibinfo {author}
  {\bibfnamefont {R.}~\bibnamefont {Ahuja}},\ }\bibfield  {title} {\bibinfo
  {title} {{Impact of Edge Structures on Interfacial Interactions and Efficient
  Visible-Light Photocatalytic Activity of Metal-Semiconductor Hybrid 2D
  Materials}},\ }\href {https://doi.org/10.1039/D0CY00420K} {\bibfield
  {journal} {\bibinfo  {journal} {Catal. Sci. Technol.}\ }\textbf {\bibinfo
  {volume} {10}},\ \bibinfo {pages} {3279} (\bibinfo {year}
  {2020})}\BibitemShut {NoStop}%
\bibitem [{\citenamefont {Cao}\ \emph {et~al.}(2015{\natexlab{b}})\citenamefont
  {Cao}, \citenamefont {Shen}, \citenamefont {Liang}, \citenamefont {Chen},\
  and\ \citenamefont {Shu}}]{CSedgetmd15}%
  \BibitemOpen
  \bibfield  {author} {\bibinfo {author} {\bibfnamefont {D.}~\bibnamefont
  {Cao}}, \bibinfo {author} {\bibfnamefont {T.}~\bibnamefont {Shen}}, \bibinfo
  {author} {\bibfnamefont {P.}~\bibnamefont {Liang}}, \bibinfo {author}
  {\bibfnamefont {X.}~\bibnamefont {Chen}},\ and\ \bibinfo {author}
  {\bibfnamefont {H.}~\bibnamefont {Shu}},\ }\bibfield  {title} {\bibinfo
  {title} {{Role of Chemical Potential in Flake Shape and Edge Properties of
  Monolayer MoS$_2$}},\ }\href {https://doi.org/10.1021/jp5097713} {\bibfield
  {journal} {\bibinfo  {journal} {J. Phys. Chem. C}\ }\textbf {\bibinfo
  {volume} {119}},\ \bibinfo {pages} {4294} (\bibinfo {year}
  {2015}{\natexlab{b}})}\BibitemShut {NoStop}%
\bibitem [{\citenamefont {Cheng}\ \emph {et~al.}(2018)\citenamefont {Cheng},
  \citenamefont {Li}, \citenamefont {Wei}, \citenamefont {Yin}, \citenamefont
  {Ho}, \citenamefont {Retamal}, \citenamefont {Mohammed},\ and\ \citenamefont
  {He}}]{CLgasdet18}%
  \BibitemOpen
  \bibfield  {author} {\bibinfo {author} {\bibfnamefont {B.}~\bibnamefont
  {Cheng}}, \bibinfo {author} {\bibfnamefont {T.-Y.}\ \bibnamefont {Li}},
  \bibinfo {author} {\bibfnamefont {P.-C.}\ \bibnamefont {Wei}}, \bibinfo
  {author} {\bibfnamefont {J.}~\bibnamefont {Yin}}, \bibinfo {author}
  {\bibfnamefont {K.-T.}\ \bibnamefont {Ho}}, \bibinfo {author} {\bibfnamefont
  {J.~R.~D.}\ \bibnamefont {Retamal}}, \bibinfo {author} {\bibfnamefont
  {O.~F.}\ \bibnamefont {Mohammed}},\ and\ \bibinfo {author} {\bibfnamefont
  {J.-H.}\ \bibnamefont {He}},\ }\bibfield  {title} {\bibinfo {title}
  {{Layer-Edge Device of Two-Dimensional Hybrid Perovskites}},\ }\href
  {https://doi.org/10.1038/s41467-018-07656-2} {\bibfield  {journal} {\bibinfo
  {journal} {Nat. Commun.}\ }\textbf {\bibinfo {volume} {9}},\ \bibinfo {pages}
  {5196} (\bibinfo {year} {2018})}\BibitemShut {NoStop}%
\bibitem [{\citenamefont {Zhang}\ \emph {et~al.}(2019)\citenamefont {Zhang},
  \citenamefont {Fang}, \citenamefont {Long},\ and\ \citenamefont
  {Prezhdo}}]{ZFedgesim19}%
  \BibitemOpen
  \bibfield  {author} {\bibinfo {author} {\bibfnamefont {Z.}~\bibnamefont
  {Zhang}}, \bibinfo {author} {\bibfnamefont {W.-H.}\ \bibnamefont {Fang}},
  \bibinfo {author} {\bibfnamefont {R.}~\bibnamefont {Long}},\ and\ \bibinfo
  {author} {\bibfnamefont {O.~V.}\ \bibnamefont {Prezhdo}},\ }\bibfield
  {title} {\bibinfo {title} {{Exciton Dissociation and Suppressed Charge
  Recombination at 2D Perovskite Edges: Key Roles of Unsaturated Halide Bonds
  and Thermal Disorder}},\ }\href {https://doi.org/10.1021/jacs.9b06046}
  {\bibfield  {journal} {\bibinfo  {journal} {J. Am. Chem. Soc.}\ }\textbf
  {\bibinfo {volume} {141}},\ \bibinfo {pages} {15557} (\bibinfo {year}
  {2019})}\BibitemShut {NoStop}%
\bibitem [{\citenamefont {Billing}\ and\ \citenamefont
  {Lemmerer}(2007)}]{BLcryst07}%
  \BibitemOpen
  \bibfield  {author} {\bibinfo {author} {\bibfnamefont {D.~G.}\ \bibnamefont
  {Billing}}\ and\ \bibinfo {author} {\bibfnamefont {A.}~\bibnamefont
  {Lemmerer}},\ }\bibfield  {title} {\bibinfo {title} {{Synthesis,
  Characterization and Phase Transitions in the Inorganic-Organic Layered
  Perovskite-Type Hybrids [(C$_{n}$H$_{2n+1}$NH$_{3}$)$_{2}$PbI$_{4}$], $n$ =
  4, 5 and 6}},\ }\href
  {https://doi.org/https://doi.org/10.1107/S0108768107031758} {\bibfield
  {journal} {\bibinfo  {journal} {Acta Crystallogr. Sect. B: Struct. Sci.,
  Cryst. Eng. Mater.}\ }\textbf {\bibinfo {volume} {63}},\ \bibinfo {pages}
  {735} (\bibinfo {year} {2007})}\BibitemShut {NoStop}%
\bibitem [{\citenamefont {Mitzi}(1996)}]{Mcryst96}%
  \BibitemOpen
  \bibfield  {author} {\bibinfo {author} {\bibfnamefont {D.~B.}\ \bibnamefont
  {Mitzi}},\ }\bibfield  {title} {\bibinfo {title} {{Synthesis, Crystal
  Structure, and Optical and Thermal Properties of (C$_4$H$_9$NH$_3$)$_2$MI$_4$
  (M = Ge, Sn, Pb)}},\ }\href {https://doi.org/10.1021/cm9505097} {\bibfield
  {journal} {\bibinfo  {journal} {Chem. Mater.}\ }\textbf {\bibinfo {volume}
  {8}},\ \bibinfo {pages} {791} (\bibinfo {year} {1996})}\BibitemShut {NoStop}%
\bibitem [{\citenamefont {Soe}\ \emph {et~al.}(2019)\citenamefont {Soe},
  \citenamefont {Nagabhushana}, \citenamefont {Shivaramaiah}, \citenamefont
  {Tsai}, \citenamefont {Nie}, \citenamefont {Blancon}, \citenamefont
  {Melkonyan}, \citenamefont {Cao}, \citenamefont {Traor{\'e}}, \citenamefont
  {Pedesseau}, \citenamefont {Kepenekian}, \citenamefont {Katan}, \citenamefont
  {Even}, \citenamefont {Marks}, \citenamefont {Navrotsky}, \citenamefont
  {Mohite}, \citenamefont {Stoumpos},\ and\ \citenamefont
  {Kanatzidis}}]{SNtilt19}%
  \BibitemOpen
  \bibfield  {author} {\bibinfo {author} {\bibfnamefont {C.~M.~M.}\
  \bibnamefont {Soe}}, \bibinfo {author} {\bibfnamefont {G.~P.}\ \bibnamefont
  {Nagabhushana}}, \bibinfo {author} {\bibfnamefont {R.}~\bibnamefont
  {Shivaramaiah}}, \bibinfo {author} {\bibfnamefont {H.}~\bibnamefont {Tsai}},
  \bibinfo {author} {\bibfnamefont {W.}~\bibnamefont {Nie}}, \bibinfo {author}
  {\bibfnamefont {J.-C.}\ \bibnamefont {Blancon}}, \bibinfo {author}
  {\bibfnamefont {F.}~\bibnamefont {Melkonyan}}, \bibinfo {author}
  {\bibfnamefont {D.~H.}\ \bibnamefont {Cao}}, \bibinfo {author} {\bibfnamefont
  {B.}~\bibnamefont {Traor{\'e}}}, \bibinfo {author} {\bibfnamefont
  {L.}~\bibnamefont {Pedesseau}}, \bibinfo {author} {\bibfnamefont
  {M.}~\bibnamefont {Kepenekian}}, \bibinfo {author} {\bibfnamefont
  {C.}~\bibnamefont {Katan}}, \bibinfo {author} {\bibfnamefont
  {J.}~\bibnamefont {Even}}, \bibinfo {author} {\bibfnamefont {T.~J.}\
  \bibnamefont {Marks}}, \bibinfo {author} {\bibfnamefont {A.}~\bibnamefont
  {Navrotsky}}, \bibinfo {author} {\bibfnamefont {A.~D.}\ \bibnamefont
  {Mohite}}, \bibinfo {author} {\bibfnamefont {C.~C.}\ \bibnamefont
  {Stoumpos}},\ and\ \bibinfo {author} {\bibfnamefont {M.~G.}\ \bibnamefont
  {Kanatzidis}},\ }\bibfield  {title} {\bibinfo {title} {{Structural and
  Thermodynamic Limits of Layer Thickness in 2D Halide Perovskites}},\ }\href
  {https://doi.org/10.1073/pnas.1811006115} {\bibfield  {journal} {\bibinfo
  {journal} {Proc. Natl. Acad. Sci. U. S. A.}\ }\textbf {\bibinfo {volume}
  {116}},\ \bibinfo {pages} {58} (\bibinfo {year} {2019})}\BibitemShut
  {NoStop}%
\bibitem [{\citenamefont {Glazer}(1972)}]{Gtilt72}%
  \BibitemOpen
  \bibfield  {author} {\bibinfo {author} {\bibfnamefont {A.~M.}\ \bibnamefont
  {Glazer}},\ }\bibfield  {title} {\bibinfo {title} {{The Classification of
  Tilted Octahedra in Perovskites}},\ }\href
  {https://doi.org/10.1107/S0567740872007976} {\bibfield  {journal} {\bibinfo
  {journal} {Acta Crystallogr. Sect. B: Struct. Sci., Cryst. Eng. Mater.}\
  }\textbf {\bibinfo {volume} {28}},\ \bibinfo {pages} {3384} (\bibinfo {year}
  {1972})}\BibitemShut {NoStop}%
\bibitem [{\citenamefont {Cammarata}(1994)}]{Cedstr94}%
  \BibitemOpen
  \bibfield  {author} {\bibinfo {author} {\bibfnamefont {R.~C.}\ \bibnamefont
  {Cammarata}},\ }\bibfield  {title} {\bibinfo {title} {{Surface and Interface
  Stress Effects in Thin Films}},\ }\href
  {https://doi.org/https://doi.org/10.1016/0079-6816(94)90005-1} {\bibfield
  {journal} {\bibinfo  {journal} {Prog. Surf. Sci.}\ }\textbf {\bibinfo
  {volume} {46}},\ \bibinfo {pages} {1} (\bibinfo {year} {1994})}\BibitemShut
  {NoStop}%
\bibitem [{\citenamefont {Falin}\ \emph {et~al.}(2017)\citenamefont {Falin},
  \citenamefont {Cai}, \citenamefont {Santos}, \citenamefont {Scullion},
  \citenamefont {Qian}, \citenamefont {Zhang}, \citenamefont {Yang},
  \citenamefont {Huang}, \citenamefont {Watanabe}, \citenamefont {Taniguchi},
  \citenamefont {Barnett}, \citenamefont {Chen}, \citenamefont {Ruoff},\ and\
  \citenamefont {Li}}]{FCgrbn17}%
  \BibitemOpen
  \bibfield  {author} {\bibinfo {author} {\bibfnamefont {A.}~\bibnamefont
  {Falin}}, \bibinfo {author} {\bibfnamefont {Q.}~\bibnamefont {Cai}}, \bibinfo
  {author} {\bibfnamefont {E.~J.~G.}\ \bibnamefont {Santos}}, \bibinfo {author}
  {\bibfnamefont {D.}~\bibnamefont {Scullion}}, \bibinfo {author}
  {\bibfnamefont {D.}~\bibnamefont {Qian}}, \bibinfo {author} {\bibfnamefont
  {R.}~\bibnamefont {Zhang}}, \bibinfo {author} {\bibfnamefont
  {Z.}~\bibnamefont {Yang}}, \bibinfo {author} {\bibfnamefont {S.}~\bibnamefont
  {Huang}}, \bibinfo {author} {\bibfnamefont {K.}~\bibnamefont {Watanabe}},
  \bibinfo {author} {\bibfnamefont {T.}~\bibnamefont {Taniguchi}}, \bibinfo
  {author} {\bibfnamefont {M.~R.}\ \bibnamefont {Barnett}}, \bibinfo {author}
  {\bibfnamefont {Y.}~\bibnamefont {Chen}}, \bibinfo {author} {\bibfnamefont
  {R.~S.}\ \bibnamefont {Ruoff}},\ and\ \bibinfo {author} {\bibfnamefont
  {L.~H.}\ \bibnamefont {Li}},\ }\bibfield  {title} {\bibinfo {title}
  {{Mechanical Properties of Atomically Thin Boron Nitride and the Role of
  Interlayer Interactions}},\ }\href@noop {} {\bibfield  {journal} {\bibinfo
  {journal} {Nat. Commun.}\ }\textbf {\bibinfo {volume} {8}},\ \bibinfo {pages}
  {15815} (\bibinfo {year} {2017})}\BibitemShut {NoStop}%
\bibitem [{\citenamefont {Wei}\ and\ \citenamefont {Peng}(2014)}]{WPphos14}%
  \BibitemOpen
  \bibfield  {author} {\bibinfo {author} {\bibfnamefont {Q.}~\bibnamefont
  {Wei}}\ and\ \bibinfo {author} {\bibfnamefont {X.}~\bibnamefont {Peng}},\
  }\bibfield  {title} {\bibinfo {title} {{Superior Mechanical Flexibility of
  Phosphorene and Few-Layer Black Phosphorus}},\ }\href
  {https://doi.org/10.1063/1.4885215} {\bibfield  {journal} {\bibinfo
  {journal} {Appl. Phys. Lett.}\ }\textbf {\bibinfo {volume} {104}},\ \bibinfo
  {pages} {251915} (\bibinfo {year} {2014})}\BibitemShut {NoStop}%
\bibitem [{\citenamefont {Bertolazzi}\ \emph {et~al.}(2011)\citenamefont
  {Bertolazzi}, \citenamefont {Brivio},\ and\ \citenamefont {Kis}}]{BBmos211}%
  \BibitemOpen
  \bibfield  {author} {\bibinfo {author} {\bibfnamefont {S.}~\bibnamefont
  {Bertolazzi}}, \bibinfo {author} {\bibfnamefont {J.}~\bibnamefont {Brivio}},\
  and\ \bibinfo {author} {\bibfnamefont {A.}~\bibnamefont {Kis}},\ }\bibfield
  {title} {\bibinfo {title} {{Stretching and Breaking of Ultrathin MoS$_2$}},\
  }\href {https://doi.org/10.1021/nn203879f} {\bibfield  {journal} {\bibinfo
  {journal} {ACS Nano}\ }\textbf {\bibinfo {volume} {5}},\ \bibinfo {pages}
  {9703} (\bibinfo {year} {2011})}\BibitemShut {NoStop}%
\bibitem [{\citenamefont {Wu}\ \emph {et~al.}(2018)\citenamefont {Wu},
  \citenamefont {Lu}, \citenamefont {Li}, \citenamefont {Sun}, \citenamefont
  {Wong},\ and\ \citenamefont {Yang}}]{WLdos18}%
  \BibitemOpen
  \bibfield  {author} {\bibinfo {author} {\bibfnamefont {L.}~\bibnamefont
  {Wu}}, \bibinfo {author} {\bibfnamefont {P.}~\bibnamefont {Lu}}, \bibinfo
  {author} {\bibfnamefont {Y.}~\bibnamefont {Li}}, \bibinfo {author}
  {\bibfnamefont {Y.}~\bibnamefont {Sun}}, \bibinfo {author} {\bibfnamefont
  {J.}~\bibnamefont {Wong}},\ and\ \bibinfo {author} {\bibfnamefont
  {K.}~\bibnamefont {Yang}},\ }\bibfield  {title} {\bibinfo {title}
  {{First-Principles Characterization of Two-Dimensional
  (CH$_{3}$(CH$_{2}$)$_{3}$NH$_{3}$)$_{2}$(CH$_{3}$NH$_{3}$)$_{n-1}$Ge$_{n}$I$_{3n+1}$
  Perovskite}},\ }\href {https://doi.org/10.1039/C8TA10055A} {\bibfield
  {journal} {\bibinfo  {journal} {J. Mater. Chem. A}\ }\textbf {\bibinfo
  {volume} {6}},\ \bibinfo {pages} {24389} (\bibinfo {year}
  {2018})}\BibitemShut {NoStop}%
\bibitem [{\citenamefont {Stoumpos}\ \emph {et~al.}(2017)\citenamefont
  {Stoumpos}, \citenamefont {Soe}, \citenamefont {Tsai}, \citenamefont {Nie},
  \citenamefont {Blancon}, \citenamefont {Cao}, \citenamefont {Liu},
  \citenamefont {Traor\'{e}}, \citenamefont {Katan}, \citenamefont {Even},
  \citenamefont {Mohite},\ and\ \citenamefont {Kanatzidis}}]{SSdos17}%
  \BibitemOpen
  \bibfield  {author} {\bibinfo {author} {\bibfnamefont {C.~C.}\ \bibnamefont
  {Stoumpos}}, \bibinfo {author} {\bibfnamefont {C.~M.~M.}\ \bibnamefont
  {Soe}}, \bibinfo {author} {\bibfnamefont {H.}~\bibnamefont {Tsai}}, \bibinfo
  {author} {\bibfnamefont {W.}~\bibnamefont {Nie}}, \bibinfo {author}
  {\bibfnamefont {J.-C.}\ \bibnamefont {Blancon}}, \bibinfo {author}
  {\bibfnamefont {D.~H.}\ \bibnamefont {Cao}}, \bibinfo {author} {\bibfnamefont
  {F.}~\bibnamefont {Liu}}, \bibinfo {author} {\bibfnamefont {B.}~\bibnamefont
  {Traor\'{e}}}, \bibinfo {author} {\bibfnamefont {C.}~\bibnamefont {Katan}},
  \bibinfo {author} {\bibfnamefont {J.}~\bibnamefont {Even}}, \bibinfo {author}
  {\bibfnamefont {A.~D.}\ \bibnamefont {Mohite}},\ and\ \bibinfo {author}
  {\bibfnamefont {M.~G.}\ \bibnamefont {Kanatzidis}},\ }\bibfield  {title}
  {\bibinfo {title} {{High Members of the 2D Ruddlesden-Popper Halide
  Perovskites: Synthesis, Optical Properties, and Solar Cells of
  (CH$_{3}$(CH$_{2}$)$_{3}$NH$_{3}$)$_{2}$(CH$_{3}$NH$_{3}$)$_{4}$Pb$_{5}$I$_{16}$}},\
  }\href {https://doi.org/https://doi.org/10.1016/j.chempr.2017.02.004}
  {\bibfield  {journal} {\bibinfo  {journal} {Chem}\ }\textbf {\bibinfo
  {volume} {2}},\ \bibinfo {pages} {427} (\bibinfo {year} {2017})}\BibitemShut
  {NoStop}%
\bibitem [{\citenamefont {Bernal}\ and\ \citenamefont {Yang}(2014)}]{BYdos14}%
  \BibitemOpen
  \bibfield  {author} {\bibinfo {author} {\bibfnamefont {C.}~\bibnamefont
  {Bernal}}\ and\ \bibinfo {author} {\bibfnamefont {K.}~\bibnamefont {Yang}},\
  }\bibfield  {title} {\bibinfo {title} {{First-Principles Hybrid Functional
  Study of the Organic-Inorganic Perovskites CH$_{3}$NH$_{3}$SnBr$_{3}$ and
  CH$_{3}$NH$_{3}$SnI$_{3}$}},\ }\href {https://doi.org/10.1021/jp509358f}
  {\bibfield  {journal} {\bibinfo  {journal} {J. Phys. Chem. C}\ }\textbf
  {\bibinfo {volume} {118}},\ \bibinfo {pages} {24383} (\bibinfo {year}
  {2014})}\BibitemShut {NoStop}%
\bibitem [{\citenamefont {Kohn}\ and\ \citenamefont {Sham}(1965)}]{KSexcorr65}%
  \BibitemOpen
  \bibfield  {author} {\bibinfo {author} {\bibfnamefont {W.}~\bibnamefont
  {Kohn}}\ and\ \bibinfo {author} {\bibfnamefont {L.~J.}\ \bibnamefont
  {Sham}},\ }\bibfield  {title} {\bibinfo {title} {{Self-Consistent Equations
  Including Exchange and Correlation Effects}},\ }\href
  {https://doi.org/10.1103/PhysRev.140.A1133} {\bibfield  {journal} {\bibinfo
  {journal} {Phys. Rev.}\ }\textbf {\bibinfo {volume} {140}},\ \bibinfo {pages}
  {A1133} (\bibinfo {year} {1965})}\BibitemShut {NoStop}%
\bibitem [{\citenamefont {Sholl}\ and\ \citenamefont
  {Steckel}(2011)}]{SSdft11}%
  \BibitemOpen
  \bibfield  {author} {\bibinfo {author} {\bibfnamefont {D.}~\bibnamefont
  {Sholl}}\ and\ \bibinfo {author} {\bibfnamefont {J.~A.}\ \bibnamefont
  {Steckel}},\ }\href@noop {} {\emph {\bibinfo {title} {{Density Functional
  Theory: A Practical Introduction}}}}\ (\bibinfo  {publisher} {John Wiley \&
  Sons, Inc., Hoboken, New Jersey},\ \bibinfo {year} {2011})\BibitemShut
  {NoStop}%
\bibitem [{\citenamefont {Kresse}\ and\ \citenamefont
  {Furthm\"uller}(1996)}]{KFvasp96}%
  \BibitemOpen
  \bibfield  {author} {\bibinfo {author} {\bibfnamefont {G.}~\bibnamefont
  {Kresse}}\ and\ \bibinfo {author} {\bibfnamefont {J.}~\bibnamefont
  {Furthm\"uller}},\ }\bibfield  {title} {\bibinfo {title} {{Efficient
  Iterative Schemes for \textit{ab Initio} Total-Energy Calculations Using a
  Plane-Wave Basis Set}},\ }\href {https://doi.org/10.1103/PhysRevB.54.11169}
  {\bibfield  {journal} {\bibinfo  {journal} {Phys. Rev. B}\ }\textbf {\bibinfo
  {volume} {54}},\ \bibinfo {pages} {11169} (\bibinfo {year}
  {1996})}\BibitemShut {NoStop}%
\bibitem [{\citenamefont {Perdew}\ \emph {et~al.}(1996)\citenamefont {Perdew},
  \citenamefont {Burke},\ and\ \citenamefont {Ernzerhof}}]{PBE96}%
  \BibitemOpen
  \bibfield  {author} {\bibinfo {author} {\bibfnamefont {J.~P.}\ \bibnamefont
  {Perdew}}, \bibinfo {author} {\bibfnamefont {K.}~\bibnamefont {Burke}},\ and\
  \bibinfo {author} {\bibfnamefont {M.}~\bibnamefont {Ernzerhof}},\ }\bibfield
  {title} {\bibinfo {title} {{Generalized Gradient Approximation Made
  Simple}},\ }\href {https://doi.org/10.1103/PhysRevLett.77.3865} {\bibfield
  {journal} {\bibinfo  {journal} {Phys. Rev. Lett.}\ }\textbf {\bibinfo
  {volume} {77}},\ \bibinfo {pages} {3865} (\bibinfo {year}
  {1996})}\BibitemShut {NoStop}%
\bibitem [{\citenamefont {Grimme}(2006)}]{DFTD206}%
  \BibitemOpen
  \bibfield  {author} {\bibinfo {author} {\bibfnamefont {S.}~\bibnamefont
  {Grimme}},\ }\bibfield  {title} {\bibinfo {title} {{Semiempirical GGA-Type
  Density Functional Constructed with a Long-Range Dispersion Correction}},\
  }\href {https://doi.org/10.1002/jcc.20495} {\bibfield  {journal} {\bibinfo
  {journal} {J. Comput. Chem.}\ }\textbf {\bibinfo {volume} {27}},\ \bibinfo
  {pages} {1787} (\bibinfo {year} {2006})}\BibitemShut {NoStop}%
\end{thebibliography}%

\end{document}